\documentclass[10pt, preprint]{aastex}
\usepackage{natbib}
\singlespace
\slugcomment{Accepted by the \emph{Astrophysical Journal}}

\shortauthors{Bennett et al.}

\shorttitle{MAP First Year Results}

\newcommand{\iMAP}         {{\sl WMAP}}

\newcommand{\lsim}         {\mbox{$_<\atop^{\sim}$}}
\newcommand{\gsim}         {\mbox{$_>\atop^{\sim}$}}

\newcommand{\ddeg}         {\mbox{${\rlap.}^\circ$}}

\newfont{\bi}{cmbxti10}

\begin{document}

\title{First Year {\sl Wilkinson Microwave Anisotropy Probe} ({\sl
WMAP}\altaffilmark{1}) Observations: 
Preliminary Maps and Basic Results}

\author{
C. L. Bennett \altaffilmark{2}, 
M. Halpern  \altaffilmark{3},
G. Hinshaw \altaffilmark{2}, 
N. Jarosik \altaffilmark{4},
A. Kogut \altaffilmark{2}, 
M. Limon \altaffilmark{2,5},
S. S. Meyer \altaffilmark{6},
L. Page \altaffilmark{4},
D. N. Spergel \altaffilmark{7},
G. S. Tucker \altaffilmark{2,5,8},
E. Wollack \altaffilmark{2},
E. L. Wright \altaffilmark{9},
C. Barnes \altaffilmark{4},
M. R. Greason \altaffilmark{10},
R. S. Hill \altaffilmark{10},
E. Komatsu \altaffilmark{7},
M. R. Nolta \altaffilmark{4},
N. Odegard \altaffilmark{10},
H. V. Peiris \altaffilmark{7},
L. Verde \altaffilmark{7},
J. L. Weiland \altaffilmark{10}
}

\altaffiltext{1}{\iMAP\ is the result of a partnership between Princeton 
                 University and NASA's Goddard Space Flight Center. Scientific 
                 guidance is provided by the \iMAP\ Science Team.}
\altaffiltext{2}{Code 685, Goddard Space Flight Center, Greenbelt, MD 20771}
\altaffiltext{3}{Dept. of Physics and Astronomy, University of British Columbia, Vancouver, BC  Canada V6T 1Z1}
\altaffiltext{4}{Dept. of Physics, Jadwin Hall, Princeton, NJ 08544}
\altaffiltext{5}{National Research Council (NRC) Fellow}
\altaffiltext{6}{Depts. of Astrophysics and Physics, EFI and CfCP, University of Chicago, Chicago, IL 60637}
\altaffiltext{7}{Dept of Astrophysical Sciences, Princeton University, Princeton, NJ 08544}
\altaffiltext{8}{Dept. of Physics, Brown University, Providence, RI 02912}
\altaffiltext{9}{UCLA Astronomy, PO Box 951562, Los Angeles, CA 90095-1562}
\altaffiltext{10}{Science Systems and Applications, Inc. (SSAI), 10210 Greenbelt Road, Suite 600 Lanham, Maryland 20706}

\email{Charles.L.Bennett@NASA.gov}

\begin{abstract}
We present full sky microwave maps in five 
frequency bands (23 to 94 GHz) from the \iMAP\ first year sky survey.  
Calibration errors are $<\!\!\!0.5$\% and the low systematic error level 
is well specified.  The cosmic microwave background (CMB) is 
separated from the foregrounds using multifrequency data.  The sky maps 
are consistent with the $7^\circ$ 
full-width at half-maximum (FWHM) {\sl Cosmic 
Background Explorer (COBE)} maps.  We report more precise, but consistent, 
dipole and quadrupole values.  

The CMB 
anisotropy obeys Gaussian statistics with 
$-58<f_{NL}<134$ (95\% CL).
The $2\le l\le 900$ anisotropy power spectrum  
is cosmic variance limited for $l<354$ with a signal-to-noise
ratio $>\!\!\!1$ per mode to $l=658$.  
The temperature-polarization 
cross-power spectrum reveals both acoustic features and
a large angle correlation from reionization.  
The optical depth of reionization is 
\ensuremath{\tau = 0.17 \pm 0.04}, 
which implies a reionization epoch of  
$t_r=180^{+220}_{-80}$ Myr (95\% CL)
after the Big Bang at a redshift of 
$z_r=20^{+10}_{-9}$ (95\% CL) 
for a range of ionization scenarios.  This early reionization
is incompatible with the presence of a significant 
warm dark matter density.

A best-fit cosmological model to the CMB and 
other measures of large scale structure works remarkably
well with only a few parameters.  The age of the best-fit universe is 
\ensuremath{t_0 = 13.7 \pm 0.2 \mbox{ Gyr}}  
old.   
Decoupling was 
\ensuremath{t_{dec} = 379^{+ 8}_{- 7} \mbox{ kyr}} 
after the Big Bang at a redshift of 
\ensuremath{z_{dec} = 1089 \pm 1}. 
The thickness of the decoupling surface was 
\ensuremath{\Delta z_{dec} = 195 \pm 2}.  
The matter density of the universe is 
\ensuremath{\Omega_mh^2 = 0.135^{+ 0.008}_{- 0.009}}, 
the baryon density is 
\ensuremath{\Omega_bh^2 = 0.0224 \pm 0.0009}, 
and the total mass-energy of the universe is
\ensuremath{\Omega_{tot} = 1.02 \pm 0.02}.
It appears that there may be progressively less fluctuation power on smaller scales, from 
\iMAP\ to fine scale CMB measurements to galaxies and finally to 
the Ly-$\alpha$ forest.  This may be accounted for with a running 
spectral index of scalar fluctuations, fit as 
\ensuremath{n_s = 0.93 \pm 0.03} 
at wavenumber $k_0 = 0.05$ Mpc$^{-1}$ ($l_{eff}\approx 700$), 
with a slope of 
\ensuremath{dn_s/d\ln{k} = -0.031^{+ 0.016}_{- 0.018}} in the best-fit 
model.  
(For \iMAP\ data alone, $n_s=0.99\pm 0.04$.)
This flat universe model is composed of 4.4\%
baryons, 22\% dark matter and 73\% dark energy.
The dark energy equation of state is limited 
to \ensuremath{w < -0.78 (95\%\mbox{\ CL})}.  

Inflation theory is supported with 
$n_s\approx 1$, 
$\Omega_{tot}\approx 1$, 
Gaussian random phases of the CMB anisotropy, 
and superhorizon fluctuations implied by the TE anticorrelations at decoupling. 
An admixture of isocurvature modes does not improve the fit.  
The tensor-to-scalar ratio is $r(k_0=0.002 \mbox{ Mpc}^{-1})<0.90$ (95\% CL).
The lack of CMB fluctuation power on the largest angular scales reported 
by {\sl COBE} and confirmed by \iMAP\ is intriguing.  
\iMAP\ continues to operate, so results will improve.  

\end{abstract}


\newpage

\section{INTRODUCTION}\label{intro}

The cosmic microwave background (CMB) radiation was first detected by
\citet{penzias/wilson:1965}.  After its discovery, a small number 
of experimentalists worked for years to better characterize the CMB as they 
searched for temperature fluctuations.  A leader of this effort, and of the \iMAP\ 
effort, was our recently deceased colleague, Professor David T. Wilkinson of 
Princeton University.  He 
was also a leading member of the {\sl Cosmic Background Explorer (COBE)} mission 
team, which accurately characterized the spectrum of the CMB \citep{mather/etal:1990,
mather/etal:1999}
and first discovered anisotropy \citep{smoot/etal:1992, bennett/etal:1992b, 
kogut/etal:1992, wright/etal:1992}.  The {\sl MAP} was recently renamed \iMAP\ 
in his honor.

The general recognition that the CMB is a 
primary tool for determining the global properties, content, and history of the 
universe has led to the tremendous interest and growth of the field.  
In addition to the characterization of the large-scale
anisotropy results from {\sl COBE} 
\citep{bennett/etal:1996, hinshaw/etal:1996a, hinshaw/etal:1996b, 
kogut/etal:1996b, kogut/etal:1996c, kogut/etal:1996d, gorski/etal:1996, wright/etal:1996b}, 
a host of experiments have measured the finer scale anisotropy
\citep{
benoit/etal:2003,           
grainge/etal:2003,          
pearson/etal:2002,          
ruhl/etal:2003,             
kuo/etal:2002,              
dawson/etal:2001,           
halverson/etal:2002,        
hanany/etal:2000,           
leitch/etal:2000,           
wilson/etal:2000,           
padin/etal:2001,            
romeo/etal:2001,            
harrison/etal:2000,         
peterson/etal:2000,         
baker/etal:1999,            
coble/etal:1999,            
dicker/etal:1999,           
miller/etal:1999,           
deoliveira-costa/etal:1998b,  
cheng/etal:1997,            
hancock/etal:1997,          
netterfield/etal:1997,      
piccirillo/etal:1997,       
tucker/etal:1997,           
gundersen/etal:1995,        
debernardis/etal:1994,      
ganga/etal:1993,            
myers/readhead/lawrence:1993, 
tucker/etal:1993}.
As a result of these tremendous efforts, the first acoustic peak of the anisotropy 
power spectrum has been unambiguously detected \citep{knox/page:2000, 
mauskopf/etal:2000, miller/etal:1999} and 
CMB observations have placed important 
constraints on cosmological models.  Recently, 
\citet{kovac/etal:2002} reported the first detection of CMB polarization arising 
from the anisotropic scattering of CMB photons at decoupling, ushering in a new
era of CMB polarization measurements.

The \iMAP\ mission was designed to advance observational 
cosmology by making full sky CMB maps with accuracy, precision, and reliability, as 
described by \citet{bennett/etal:2003}.  The instrument observes the temperature 
difference between two directions (as did {\sl COBE}) using two nearly identical sets of 
optics \citep{page/etal:2003, page/etal:2003b}.  These optics 
focus radiation into horns \citep{barnes/etal:2002} 
that feed differential microwave radiometers \citep{jarosik/etal:2003}.  
We produce full sky maps in five frequency bands from the 
radiometer data of temperature differences measured over the full sky.  
A CMB map 
is the most compact representation of CMB anisotropy without loss of information.

In this paper we present the maps, their properties, and a synopsis of 
the basic results of the first-year of observations.  In 
\S\ref{obs} we give a brief overview of the \iMAP\ mission.  In 
\S\ref{calsys} we summarize the data analysis, calibration, and systematic 
errors of the experiment, which are discussed in much greater detail in the 
companion papers by 
\citet{hinshaw/etal:2003b}, \citet{page/etal:2003b}, \citet{jarosik/etal:2003b}, 
and \citet{barnes/etal:2003}.  In 
\S\ref{maps} we present the maps and their sampling properties, and 
we compare the \iMAP\ and {\sl COBE} maps.  In 
\S\ref{galaxy} we summarize the foreground analyses of \citet{bennett/etal:2003}.  In 
\S\ref{gauss} we establish the Gaussian nature of the \iMAP\ anisotropy, determined in 
the companion paper of \citet{komatsu/etal:2003}. In 
\S{\ref{multipoles}} we present the dipole and quadrupole moments and 
summarize analyses of the angular power spectrum 
\citep{hinshaw/etal:2003, verde/etal:2003}. In
\S\ref{reion} we highlight the \iMAP\ polarization results, including a detection of the 
reionization of the universe \citep{kogut/etal:2003}.  In 
\S\ref{cosmo} we summarize some of the cosmological implications of the \iMAP\ 
results \citep{page/etal:2003, spergel/etal:2003, peiris/etal:2003}.  Finally, in 
\S\ref{data} we discuss the availability of the \iMAP\ data products.

\section{OBSERVATIONS \label{obs}}

The 840 kg \iMAP\ Observatory was launched aboard a Delta II 7425-10 rocket (Delta launch 
number 286) on 30 June 2001 
at 3:46:46.183 EDT from Cape Canaveral.  \iMAP\ executed three phasing loops in the Earth-Moon 
system before a lunar-gravity-assist swing-by, a month after launch, catapulted \iMAP\ to an
orbit about the second Lagrange point of the Sun-Earth system, L$_2$.  Station-keeping is 
performed approximately four times per year to maintain the Observatory in a 
Lissajous orbit about the L$_2$ point with the Earth-\iMAP\ vector within about 
$\sim 1^\circ - 10^\circ$ of the Sun-Earth vector. 
The phasing loop maneuvers and station-keeping are executed using the \iMAP\ 
propulsion system of blow-down hydrazine and eight thrusters.

The central design philosophy of the \iMAP\ mission was to minimize sources of systematic 
measurement errors \citep{bennett/etal:2003}.  The {\sl COBE} mission 
proved the effectiveness of 
a differential design in minimizing systematic errors.  Therefore, the \iMAP\ instrument 
was designed with a back-to-back optical system with 1.4 m $\times$ 1.6 m 
primary reflectors to provide for differential measurements of the sky.  
The primary and secondary reflectors direct radiation into two focal planes,
with ten feed horns in each, as described by \cite{page/etal:2003}.  

The beams have a gain pattern, $G$, which is neither symmetric nor Gaussian.  
We define the beam solid angle as $\int G(\Omega)/G_{\rm max}\; d\Omega$.
The beam size can be expressed as the
square root of the beam solid angles, giving 0.22$^\circ$, 0.35$^\circ$, 0.51$^\circ$, 
0.66$^\circ$, and 0.88$^\circ$ for W-band though K-band, respectively.  
Alternately, the beams can be expressed in terms of a full-width at half-maximum (FWHM) 
for each band, given in Table \ref{overview}.  Detailed analyses of the \iMAP\ beams
are discussed by \citet{page/etal:2003b, page/etal:2003}.

\begin{deluxetable}{llllll}
\tablecaption{Approximate Observational Properties by Band\label{overview}}
\tablewidth{0pt}
\tablehead{
\colhead{Item} & \colhead{K-Band}   & \colhead{Ka-Band}   & \colhead{Q-Band} &
\colhead{V-Band}  & \colhead{W-Band}}
\startdata
Wavelength, $\lambda$ (mm)     &  13   &  9.1  &  7.3  &  4.9  &  3.2 \\
Frequency, $\nu$ (GHz)         &  22.8   &  33.0   &   40.7  &   60.8  &  93.5  \\
Ant./therm. conversion factor, 
$\Delta{\rm T}/\Delta {\rm T_A}$       & 1.014 & 1.029 & 1.044 & 1.100 & 1.251\\
Noise, $\sigma_0$ (mK) $\sigma=\sigma_0 N_{obs}^{-1/2}$
                               & 1.424 & 1.449 & 2.211 & 3.112 &  6.498 \\
Beam width $\theta$($^\circ$FWHM)         & 0.82  &  0.62 &  0.49 &  0.33 &  0.21\\
No. of Differencing Assemblies &   1   &   1   &   2   &   2   &   4  \\
No. of Radiometers             &   2   &   2   &   4   &   4   &   8  \\
No. of Channels                &   4   &   4   &   8   &   8   &  16  \\
\enddata
\end{deluxetable}

\begin{deluxetable}{lccccc}
\tablecaption{Data Flagging Summary  \label{baddata}}
\tablewidth{0pt}
\tablehead{
\colhead{Category} & \colhead{K-Band}   & \colhead{Ka-Band}   & \colhead{Q-Band} &
\colhead{V-Band}  & \colhead{W-Band}}
\startdata
\sidehead{Rejected or Lost Data}
Lost Data (\%)                 & 0.27 & 0.27 & 0.27 & 0.27 & 0.27 \\
Spacecraft thermal change (\%) & 0.87 & 0.87 & 0.87 & 0.87 & 0.87 \\
Gain or baseline step (\%)     & 0.00 & 0.13 & 0.12 & 0.00 & 0.22 \\
                               &------&------&------&------&------\\
 Total Lost or Bad Data (\%)   & 1.04 & 1.27 & 1.26 & 1.14 & 1.36 \\
\sidehead{Data Not Used in Maps}
Planet flag (\%)               & 0.11 & 0.11 & 0.11 & 0.11 & 0.11 \\
\enddata
\end{deluxetable}

The feed horns are attached to orthomode transducers (OMTs) that split 
the polarization of the incoming signal into a differential correlation radiometer system 
with High Electron Mobility Transistor (HEMT) amplifiers.  
There are ten ``differencing assemblies'' each 
consisting of two ``radiometers'' with two ``channels'' each \citep{jarosik/etal:2003,
bennett/etal:2003}.  There are four W-band ($\sim 94 $ GHz), two V-band ($\sim$61 GHz), 
two Q-band ($\sim$ 41 GHz), one Ka-band ($\sim$ 33 GHz), and one K-band ($\sim$ 23 GHz) 
differencing assemblies.  We usually refer to these bands by 
the generic designations K, Ka, Q, V, and W because there are multiple radiometers 
in each band, whose precise frequencies are not identical.  Also, the effective frequency of a
radiometer depends on the spectrum of the emission it detects.  Precise frequencies 
for the radiometers for a CMB anisotropy spectrum are given by 
\citet{jarosik/etal:2003}.  Polynomials are given to determine the effective frequency 
of the radiometers depending on the emission frequency spectrum.  See Table \ref{overview} 
for a summary of radiometer properties.

Undesirable $1/f$ noise is minimized by the design of the \iMAP\ radiometers 
\citep{jarosik/etal:2003}.  All radiometers have $1/f$ knees below 50 mHz; 18 of 20 are 
below 10 mHz; and 10 of the 20 are below 1 mHz \cite{jarosik/etal:2003b}.  (The $1/f$ knee
is defined as the frequency where the noise power spectral density is $\sqrt 2$ times 
higher than its high frequency value.)  \citet{jarosik/etal:2003} demonstrate that all 
radiometer outputs have Gaussian noise, which ``integrates down'' with time as expected.

The radiometers are passively cooled to $\sim 90$ K with no mechanical refrigerators.  
In addition, no actively 
cycling heaters were permitted anywhere on the \iMAP\ spacecraft.  These design features 
helped to ensure a mechanically, thermally, and electronically quiet platform that minimizes 
the driving forces of systematic measurement errors.

In addition to the differential design, the {\sl COBE} mission also demonstrated the
importance of scanning large areas of the sky in a short period of time with a complex 
scan pattern.  \iMAP\ follows the {\sl COBE} example with a three-axis (three reaction wheel)  
control system that maintains the Observatory in a nearly 
constant survey mode of operations.  (The Observatory is in constant survey mode, except 
for only $\sim 1$ hr for each of $\sim 4$ station-keeping maneuvers per year.)
In survey mode, the optical boresight sweeps out a complex pattern 
on the sky \citep{bennett/etal:2003}.  Approximately 30\% of the sky is observed each hour.  
The Observatory spins at 0.464 rpm ($\sim$7.57 mHz) and precesses at 1 rev hr$^{-1}$
($\sim$ 0.3 mHz).  

Six months are required for L$_2$ to orbit half way around the Sun, allowing for full sky
coverage.  The observations presented in this and companion papers include 
a full orbit about the Sun, thus
containing two sets of full sky observations.  By 10 August 2001, \iMAP\ was sufficiently 
stable in its L$_2$ orbit for CMB data-taking to commence.  One year of observations, completed 
on 9 August 2002, were analyzed.   Data taken beyond this date will be the 
subject of future analyses.

\section{DATA PROCESSING, CALIBRATION, \& SYSTEMATIC ERRORS \label{calsys}}

Time-ordered telemetry data from the Observatory are down-linked via NASA's Deep Space Network 
(DSN) to the \iMAP\ Science and Mission Operations Center (SMOC) 
at the Goddard Space Flight Center.  The 
data are then transferred to the \iMAP\ Science Team for analysis. All of the instrument 
data are 
down-linked to the ground without any on-board flight data processing, thus allowing full insight
into potential systematic effects.

Only a fraction of a percent of data was lost in the flow from the Observatory to the 
SMOC.  About 1\% of the received data were not 
used due to systematic error concerns (e.g., data taken during near station-keeping maneuvers).  
Of the $\sim 99$\% good data, the processing pipeline 
flagged observations where bright planets were in the beams so that these data 
would not used be used in making maps.  The statistics on lost, bad, and flagged 
data are given in Table \ref{baddata}.

\begin{figure}
\figurenum{1}
\epsscale{.9}
\plotone{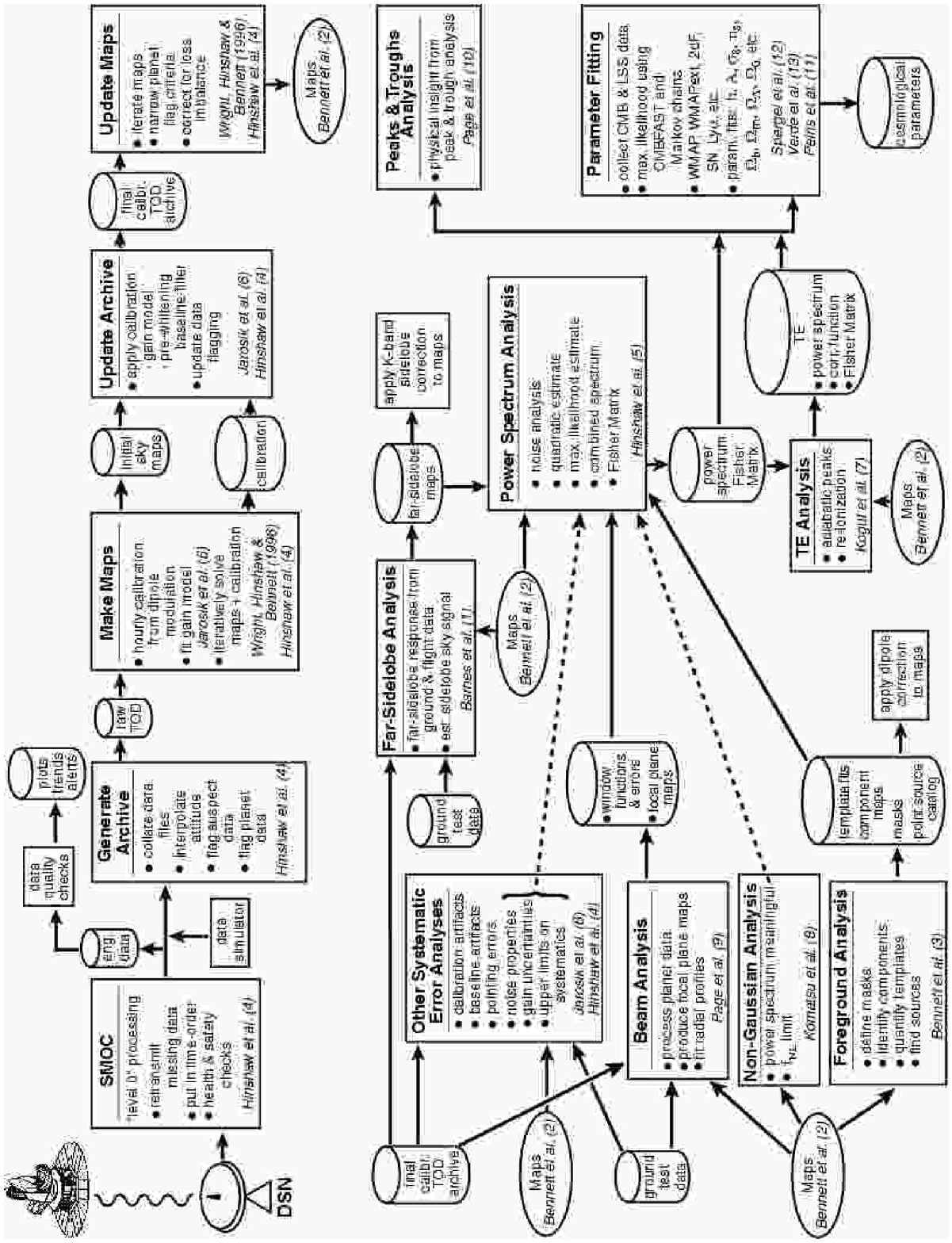}
\caption{An overview of the \iMAP\ data flow. The references are: 
(1) \citet{barnes/etal:2003}
(2) \citet{bennett/etal:2003b}
(3) \citet{bennett/etal:2003c}
(4) \citet{hinshaw/etal:2003}
(5) \citet{hinshaw/etal:2003b}
(6) \citet{jarosik/etal:2003b}
(7) \citet{kogut/etal:2003}
(8) \citet{komatsu/etal:2003}
(9) \citet{page/etal:2003b}
(10) \cite{page/etal:2003c}
(11) \citet{peiris/etal:2003}
(12) \citet{verde/etal:2003}. 
\label{dataflow}}
\end{figure}

An overview of the data flow is shown in Figure \ref{dataflow}.  
The heart of the data analysis efforts center on studies of systematic measurement errors 
\citep{hinshaw/etal:2003}.  
Components of spurious signals at the spin period are the most difficult to distinguish 
from true sky signals.  The Observatory was designed to minimize all 
thermal and voltage variations and all susceptibilities to these variations, especially 
at the spin period, as discussed in \S\ref{obs} and by \citet{bennett/etal:2003}.  
In addition, high precision temperature monitors on the Observatory provide the data 
needed to verify that systematic errors from thermal variations are negligible.
\cite{jarosik/etal:2003b} report that in-flight spin-synchronous effects from the 
radiometers are $<0.17\; \mu$K rms in the time-ordered-data (TOD), based on flight thermal 
variations multiplied by upper limits on component susceptibilities measured in ground 
testing.  Analysis of flight data without use of characterizations derived from ground-based 
testing give $<0.14\; \mu$K rms from all sources (not just the radiometers).  This is 
a factor of $>50$ times smaller than the requirement that was set in the mission's 
systematic error budget.  
Thus, {\it no corrections to the first year \iMAP\ data are required for spin-synchronous 
systematic errors.}

The core of the processing pipeline calibrates the data and converts the differential 
temperatures into maps.
The data are calibrated based on the Earth-velocity modulation of the CMB dipole.  
A gain model of the radiometers was derived and fit by \citet{jarosik/etal:2003b}.  The model is
based on the constancy of the dipole signal on the sky, the measured physical temperature 
of the front-end radiometer components, and on the time-averaged RF-bias (total power) of 
the radiometer outputs.  This relatively simple model closely matches the gains 
derived from the hourly measurements of the amplitude of the dipole and is used in 
\iMAP\ data processing.  Calibration is achieved within 0.5\% accuracy, dominated 
by the statistical
uncertainty in the absolute calibration.  

Low levels of $1/f$ noise create stripes in the maps that affect the angular 
power spectrum and other statistics derived from the maps.  A pre-whitening filter 
is applied to the TOD to minimize these artifacts.  An estimate of the magnitude of the 
striping is given in \citet{hinshaw/etal:2003} for the maps, and by  
\citet{hinshaw/etal:2003b} for the power spectrum.

The differential temperature data are formed into maps based on the technique introduced by
\citet{wright/hinshaw/bennett:1996a}.  
HEALPix\footnotemark
\footnotetext{http://www.eso.org/science/healpix/} 
is used to define map pixels on the sky in Galactic coordinates.  
Various levels of resolution are specified by a ``resolution level'' with an integer 
($r=0, 1, 2, ...$).  With 
$N_{\rm side}=2^r$, the number of pixels in the map is $N_{\rm pix}=12\;N_{\rm side}^2$.
The area per pixel is $\Omega_{\rm pix}=4\pi/N_{\rm pix}$ and the separation between pixel 
centers is $\theta_{\rm pix}=\Omega_{\rm pix}^{1/2}$.  
For example, HEALPix resolution level 
$r=9$ (used in \iMAP\ map-making) corresponds with $N_{\rm side}=512$, 
$N_{\rm pix}=3\,145\,728$, $\Omega_{\rm pix}=3.99\times 10^{-6}$ sr, 
and $\theta_{\rm pix}=0\ddeg 115 = 6.87$ arc-min.

\iMAP\ observes the sky convolved with the beam pattern.  This is equivalent to the  
the spatial transform of the sky multiplied by the instrument's ``window function.''
The beam patterns are measured in-flight from observations of Jupiter
\citep{page/etal:2003b}.  Uncertainties in our knowledge of the beam pattern, although small, 
are a significant source of uncertainty for \iMAP\ since they imply imperfect knowledge of the 
window function.  A small difference between the 
A-side and B-side optical losses was derived based on dipole observations  
and corrected in the processing.  Far sidelobes of the beam patterns, 
determined by ground measurements 
and in-flight using the Moon, have been carefully examined \citep{barnes/etal:2003}.
A small far-sidelobe correction is applied only to the K-band map.  We now describe the maps.

\begin{figure}
\figurenum{2a}
\epsscale{0.68}
\plotone{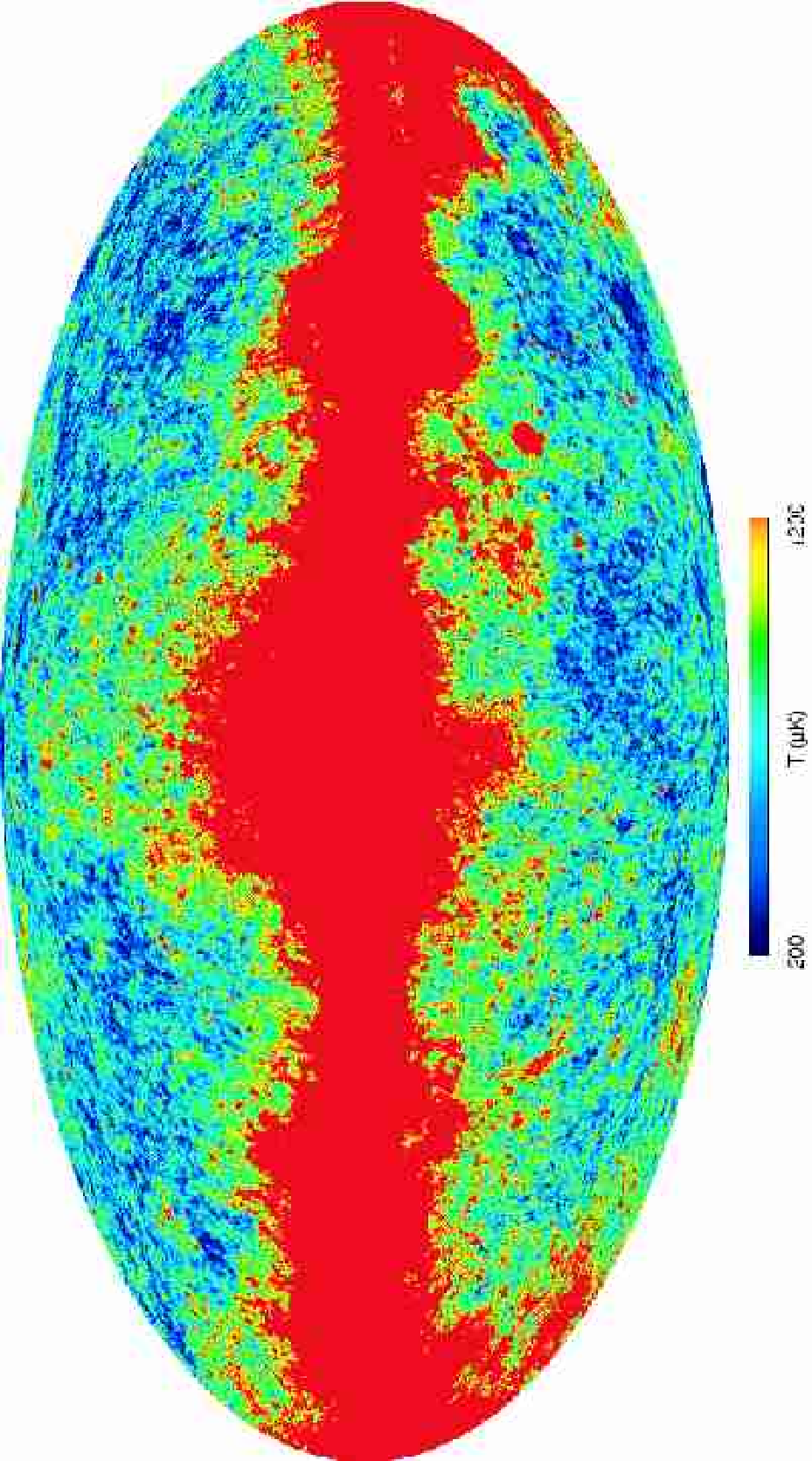}
\caption{\label{K} \iMAP\ K-band sky map in Mollweide projection 
in Galactic coordinates.} 
\end{figure}

\begin{figure}
\figurenum{2b}
\epsscale{0.68}
\plotone{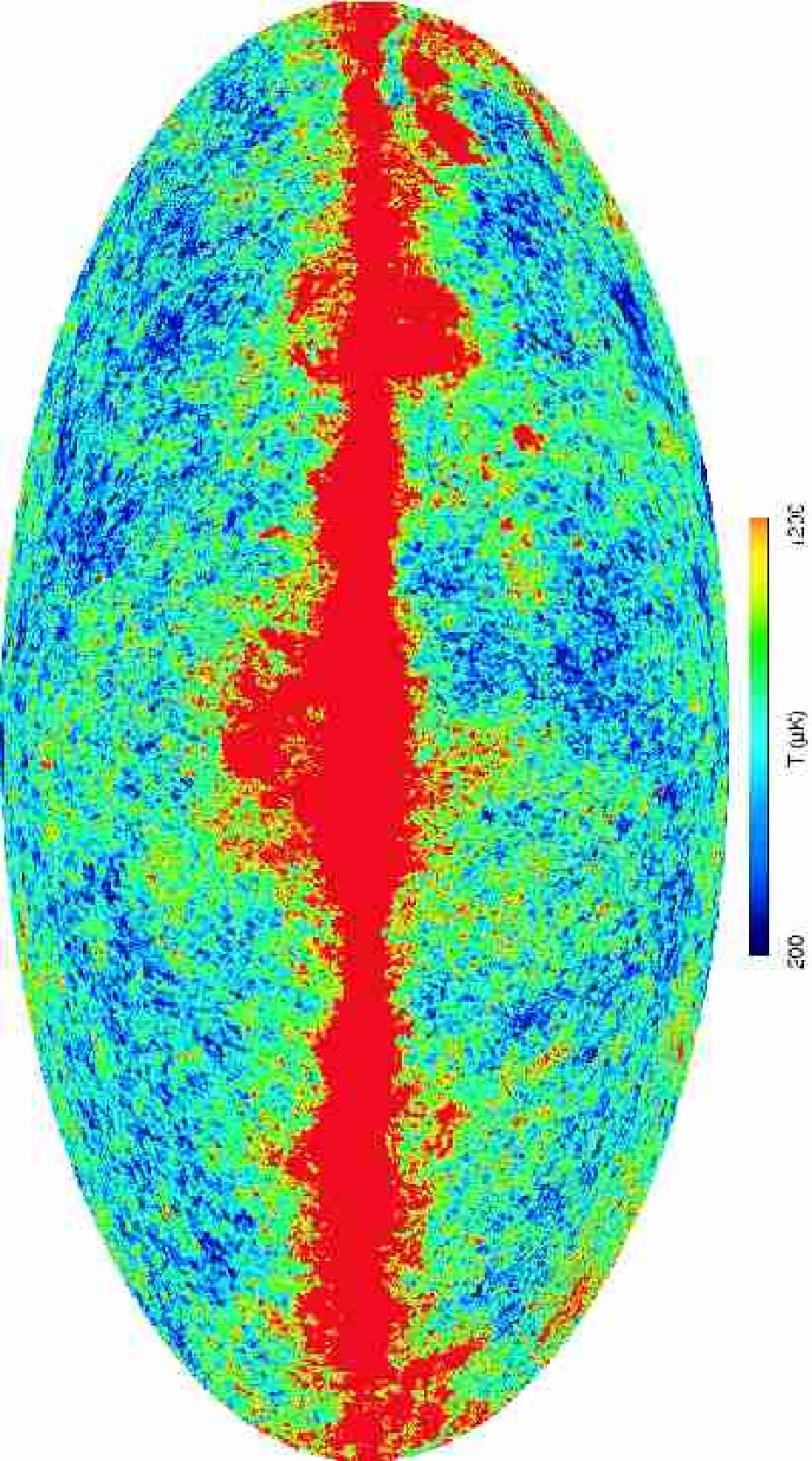}
\caption{\label{Ka} \iMAP\ Ka-band sky map in Mollweide projection 
in Galactic coordinates.} 
\end{figure}

\begin{figure}
\figurenum{2c}
\epsscale{0.68}
\plotone{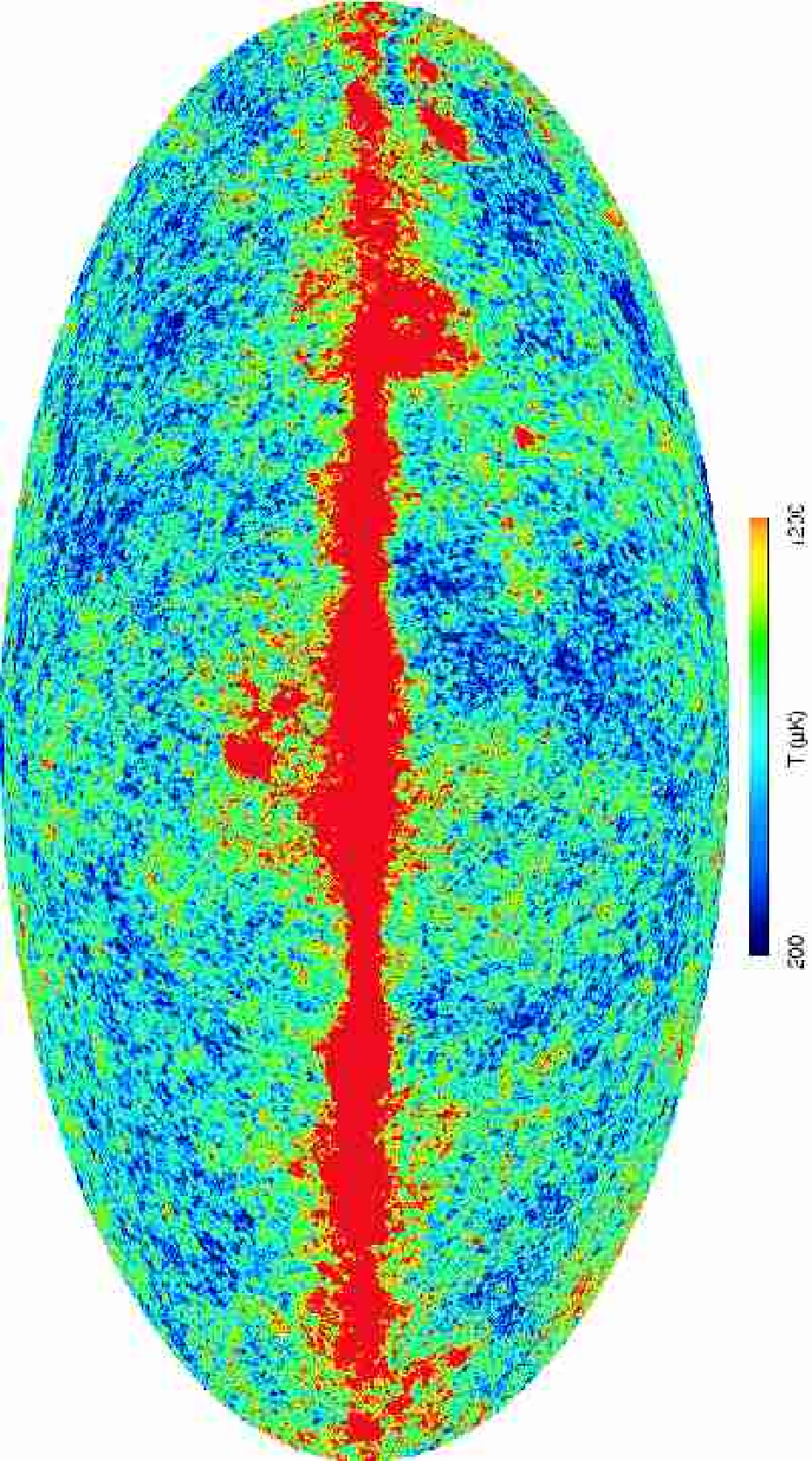}
\caption{\label{Q} \iMAP\ Q-band sky map in Mollweide projection 
in Galactic coordinates.} 
\end{figure}

\begin{figure}
\figurenum{2d}
\epsscale{0.68}
\plotone{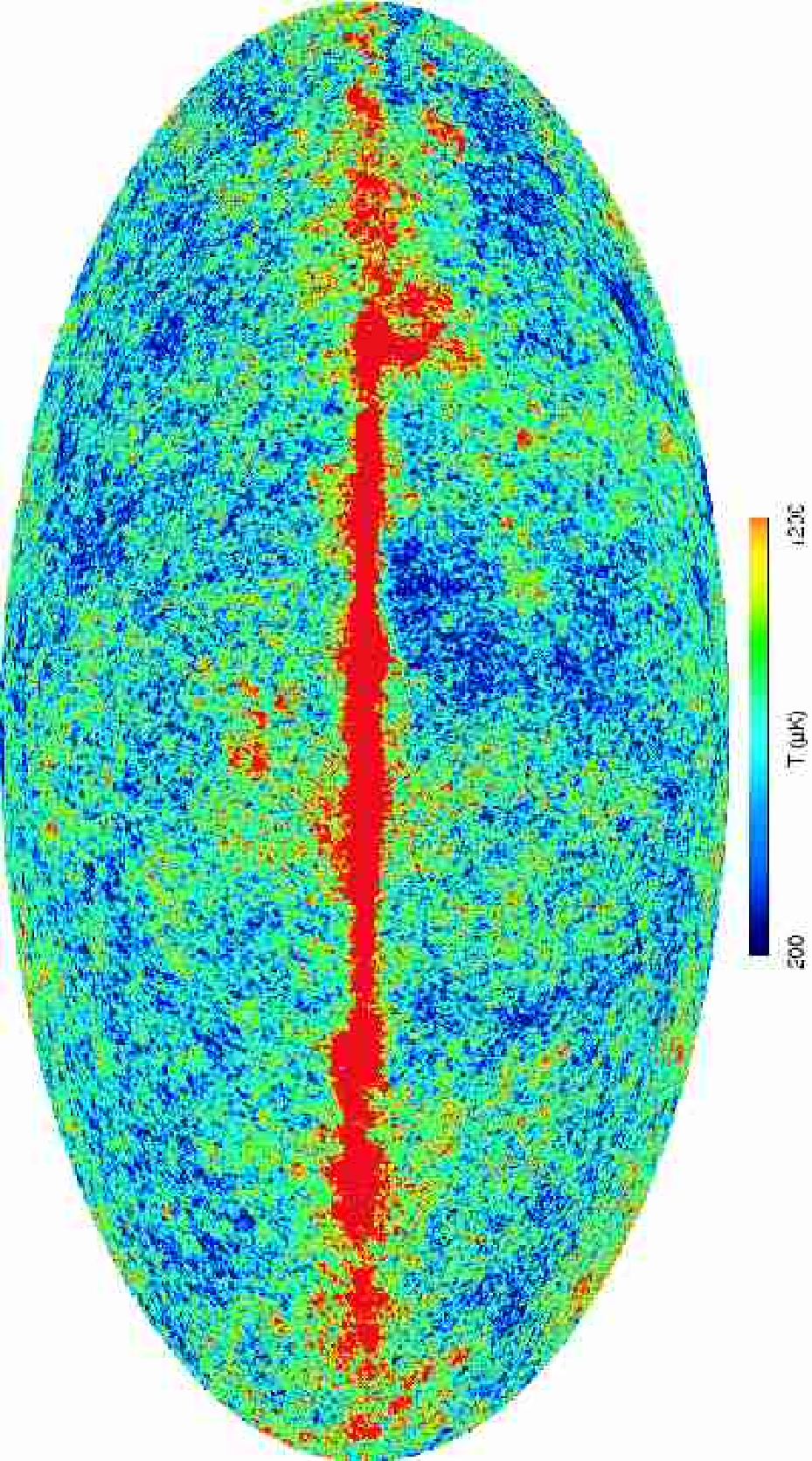}
\caption{\label{V} \iMAP\ V-band sky map in Mollweide projection 
in Galactic coordinates.} 
\end{figure}

\begin{figure}
\figurenum{2e}
\epsscale{0.68}
\plotone{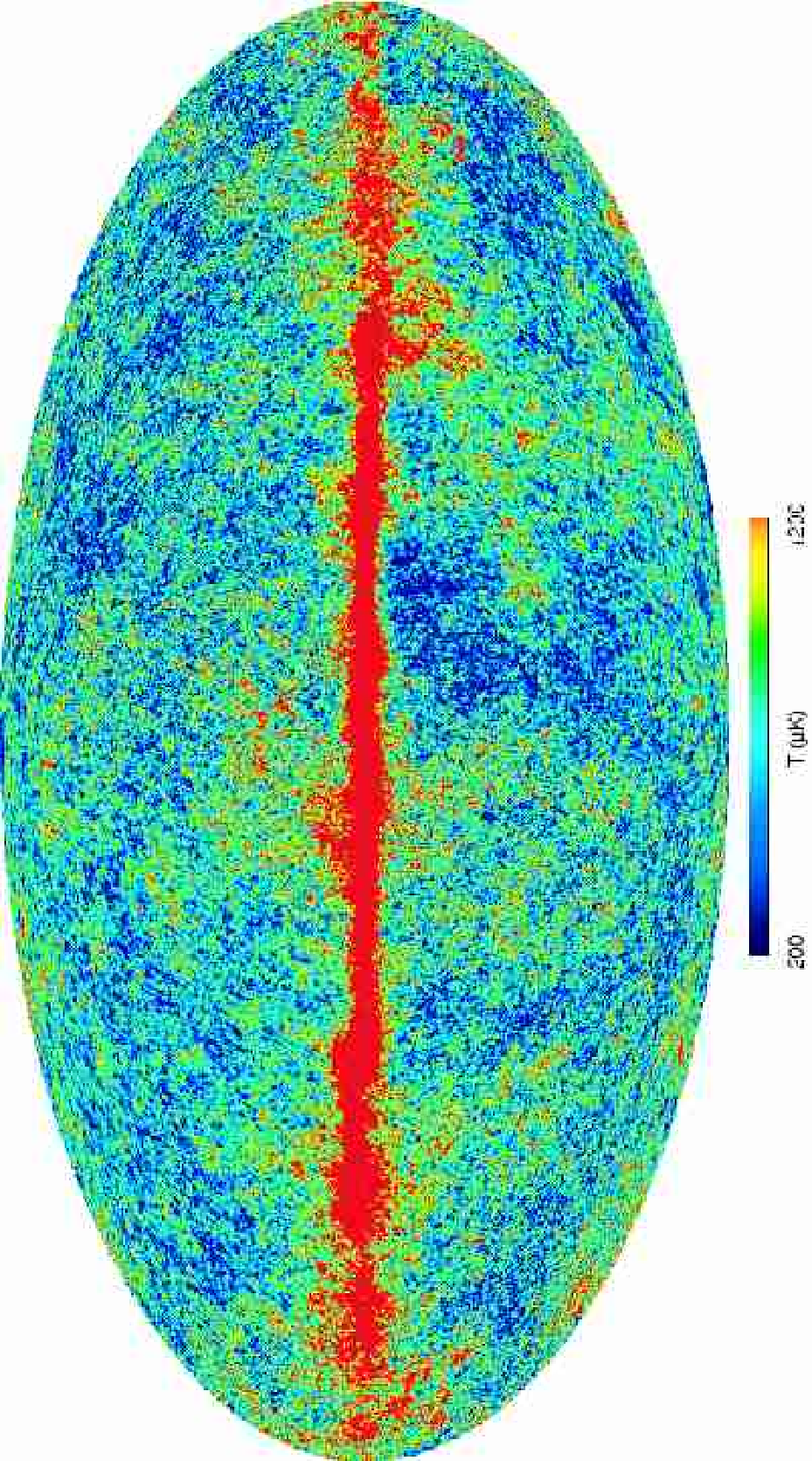}
\caption{\label{W} \iMAP\ W-band sky map in Mollweide projection 
in Galactic coordinates.} 
\end{figure}

\section{THE MAPS \label{maps}}

\begin{figure}
\figurenum{3}
\epsscale{0.982}
\plotone{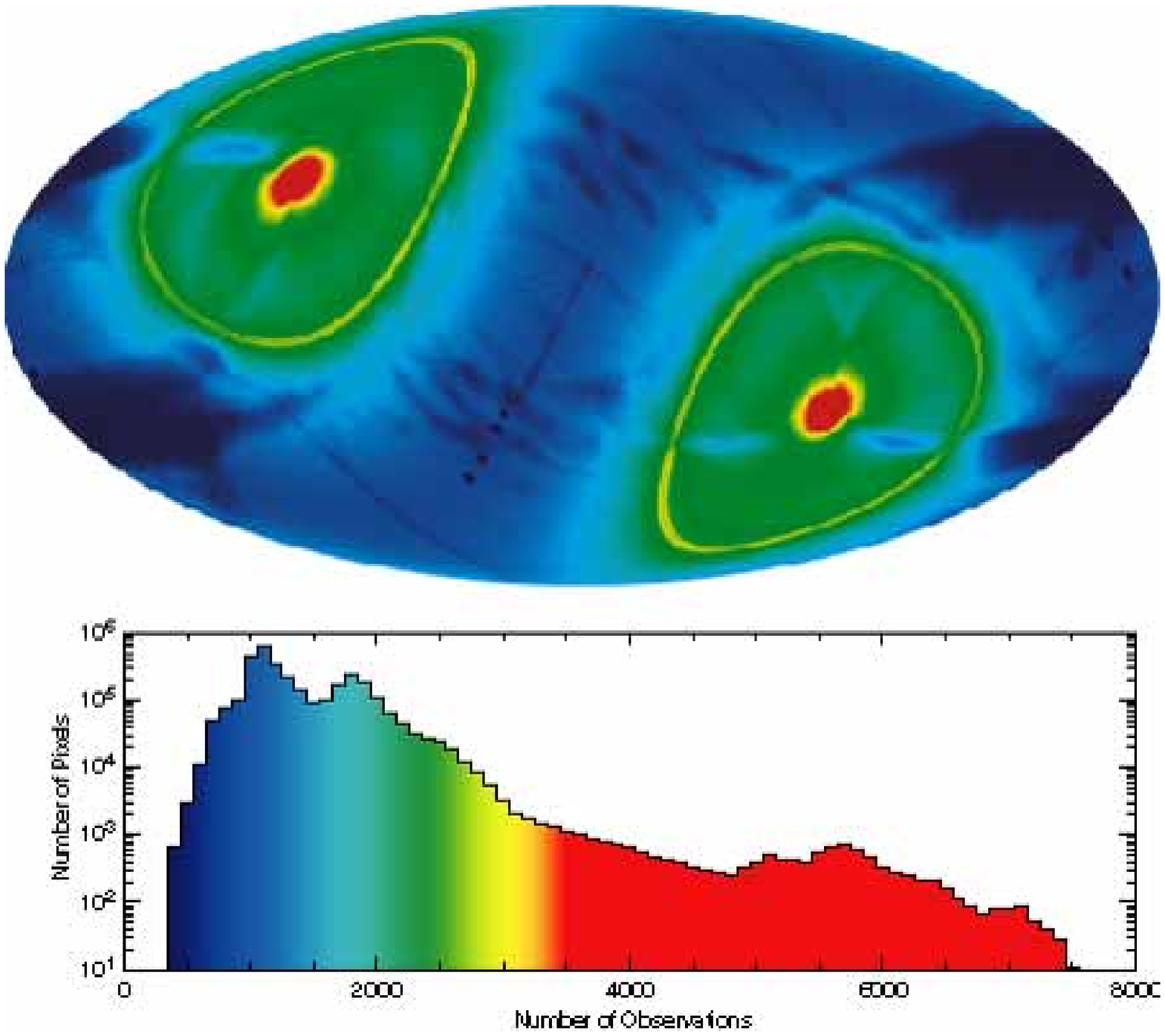}
\caption{The number of independent observations per pixel in Galactic coordinates.  
The number of observations is greatest 
at the ecliptic poles and in rings around the ecliptic poles with diameters corresponding to the 
separation angle of the two 
optical boresight directions (approximately $141^\circ$). The observations are the most sparse 
in the ecliptic plane. Small area cuts are apparent where Mars, Saturn, Jupiter, Uranus, and 
Neptune data are masked so as not to contaminate CMB analyses.  Jupiter data are used for beam 
mapping.  The histogram of the sky sampling shows the departures from 
uniform sky coverage. 
\label{Nobs_hist}
\label{Nobs}} 
\end{figure}

\begin{figure}
\figurenum{4}
\epsscale{0.618182}
\plotone{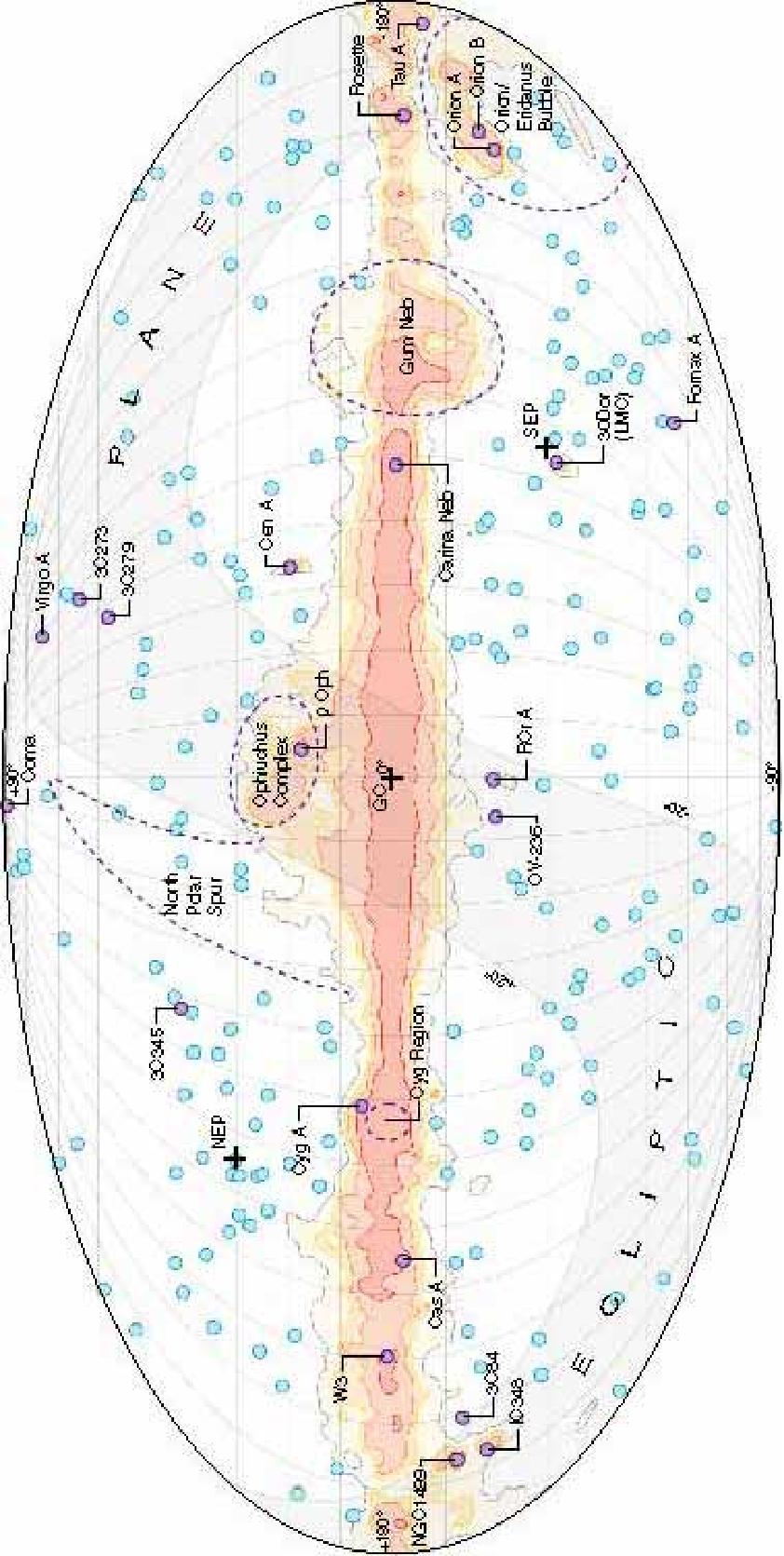}
\caption{A guide to the microwave sky for reference.  This picture shows the large-scale emission
from the Milky Way galaxy, including some of its notable components such as the Cygnus complex, the
North Polar Spur, the Gum region, etc.  The small circles show positions of the microwave
point sources detected by \iMAP\ \citep{bennett/etal:2003c}.  
The brighter sources are labeled for reference. \label{overlay}} 
\end{figure}

We combine the radiometer results within each band and present the five full sky maps
at effective CMB anisotropy frequencies 23, 33, 41, 61, and 94 GHz in Figures \ref{K}, \ref{Ka},
\ref{Q}, \ref{V}, \ref{W}.  The maps are shown in the Mollweide projection in units of CMB 
thermodynamic temperature.
The number of independent observations that contribute to each pixel 
form the sky pattern in Figure \ref{Nobs}.  
Figure \ref{overlay} provides an overall guide to some of 
the more prominent features of the maps as well as point sources detected by {\sl WMAP}, as 
described by \citet{bennett/etal:2003b}.

\begin{figure}
\figurenum{5}
\epsscale{0.982}
\plotone{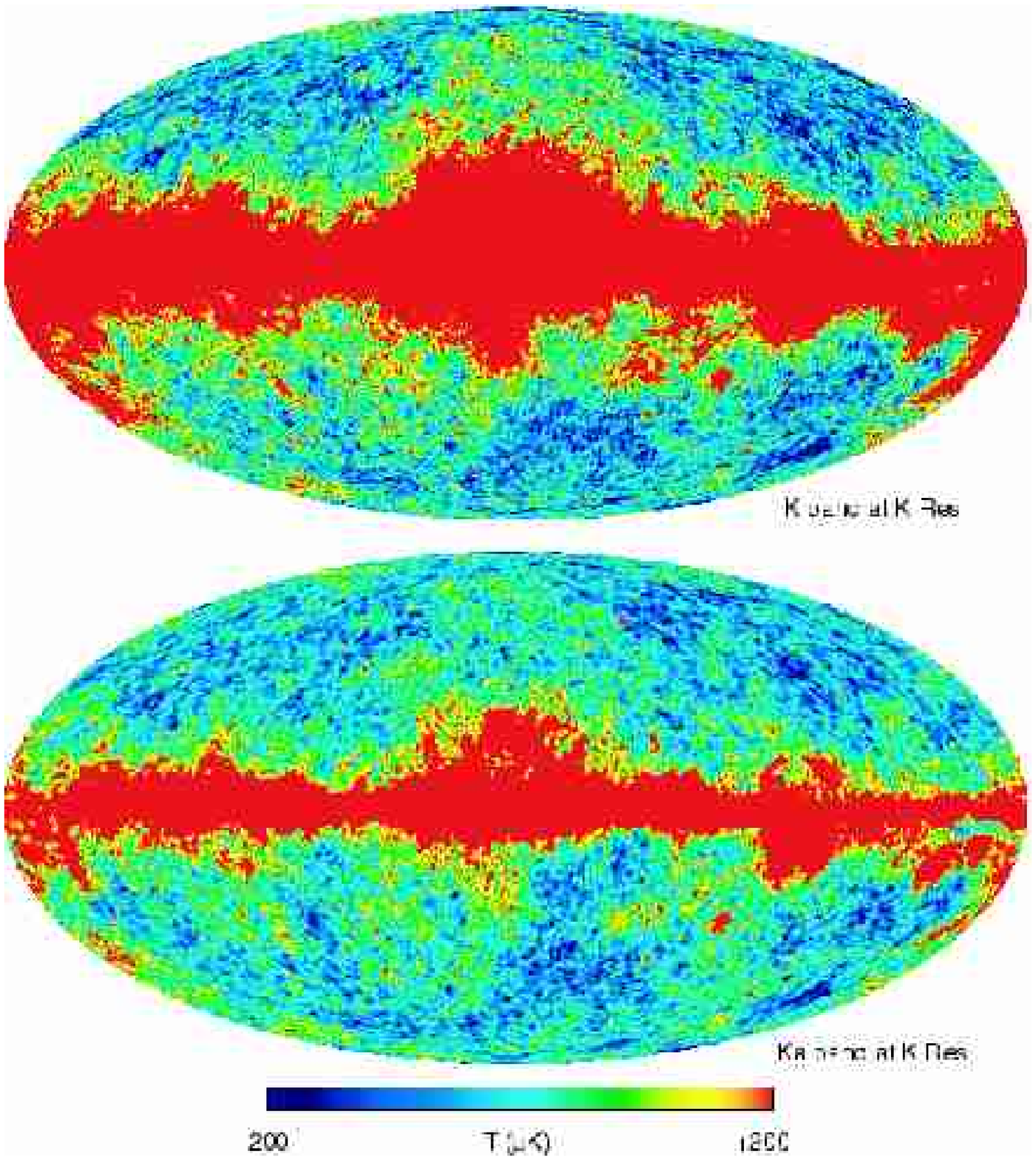}
\caption{A comparison of the K-band map with the Ka-band map smoothed to K-band resolution, both 
in thermodynamic temperature, shows the dramatic reduction of Galactic contamination with 
increased frequency.  The comparison also shows the similarity of the CMB fluctuation features 
at high Galactic latitude. \label{kka}} 
\end{figure}

\begin{figure}
\figurenum{6}
\epsscale{0.75}
\plotone{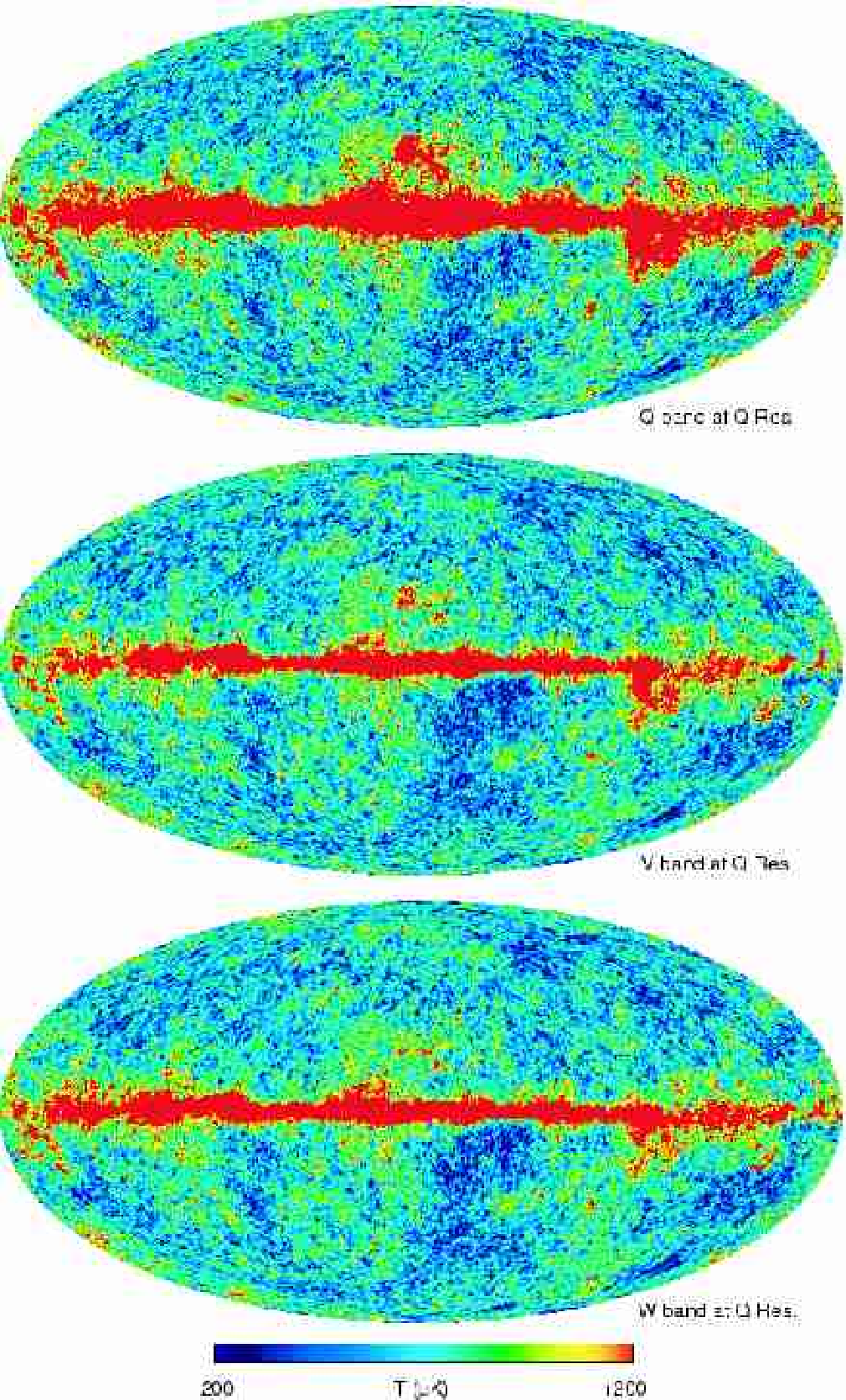}
\caption{A comparison of the Q-band, V-band, and W-band maps.  All three maps are smoothed to 
Q-band resolution and are in thermodynamic temperature.  The reduction of Galactic contamination 
relative to K-band and Ka-band (Figure \ref{kka}) is apparent.  The maps show that the constant 
features across bands are CMB anisotropy while the thermodynamic temperature of the foregrounds 
depends on the band (frequency). \label{qvw}}
\end{figure}

Figure \ref{kka} shows the K-band and Ka-band maps, with the Ka-band map smoothed to K-band 
resolution.  Note both the significant decrease in Galactic signal from K-band to Ka-band and 
the high Galactic latitude similarities of the CMB between the maps.  Likewise, Figure 
\ref{qvw} shows the Q-band, V-band, and W-band maps with the latter two smoothed to 
Q-band resolution.  Higher Galactic contamination in Q-band is apparent.  Both 
Figures \ref{kka} and \ref{qvw} highlight the consistency of the 
high Galactic latitude CMB anisotropy pattern from band to band.

\begin{figure}
\figurenum{7}
\epsscale{0.982}
\plotone{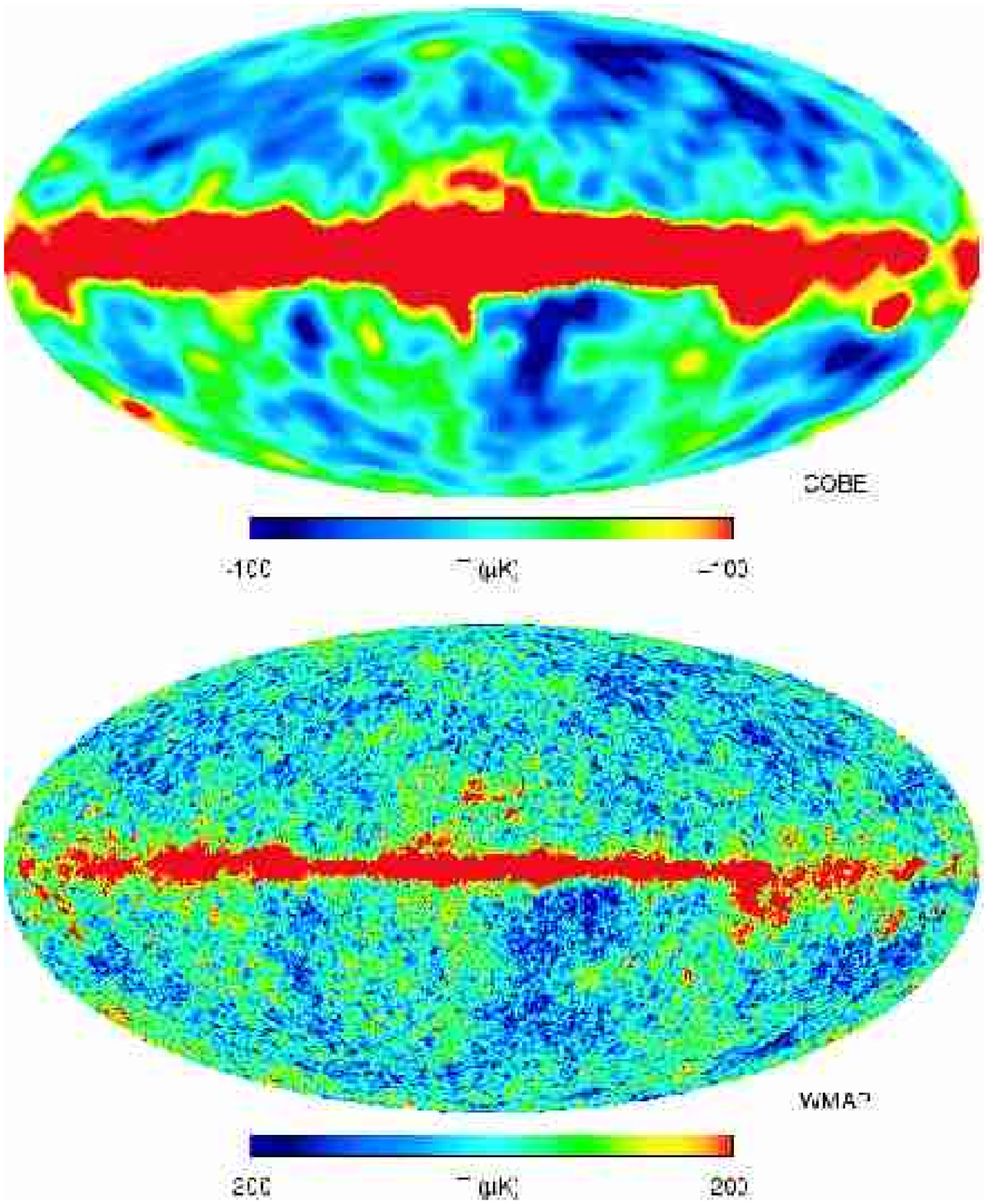}
\caption{A comparison of the {\sl COBE} 53 GHz map \citep{bennett/etal:1996} with the W-band
\iMAP\ map.  The \iMAP\ map has 30 times finer resolution than the {\sl COBE} map.\label{cobemapw}} 
\end{figure}

\begin{figure}
\figurenum{8}
\epsscale{0.982}
\plotone{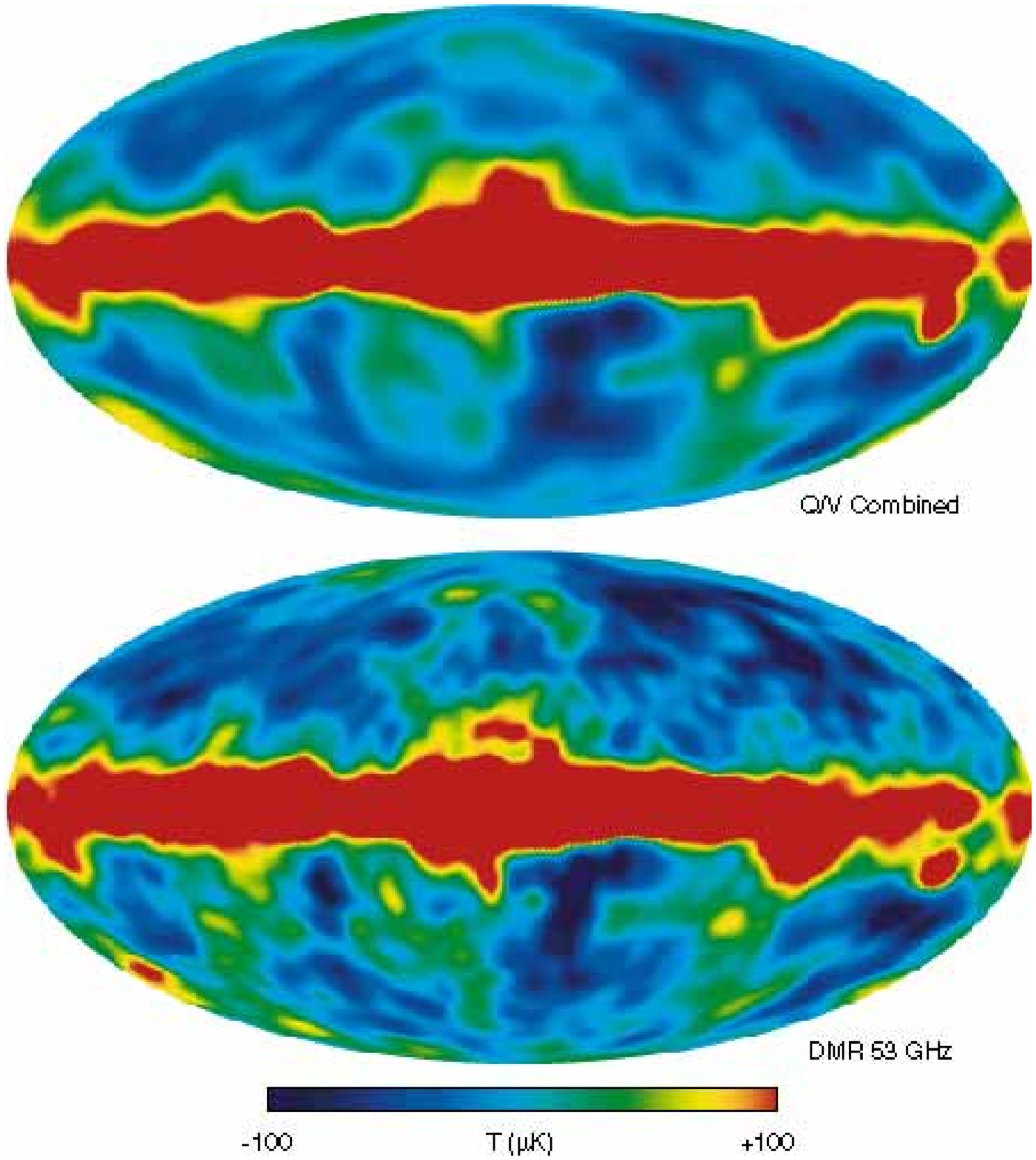}
\caption{The {\sl COBE}-DMR 53 GHz map \citep{bennett/etal:1996} is shown along with a map 
made with a linear combination of the Q-band and V-band \iMAP\ maps to mimic a 53 GHz map.  
Note the strong similarity of the maps.\label{qv53}} 
\end{figure}

\begin{figure}
\figurenum{9}
\epsscale{0.982}
\plotone{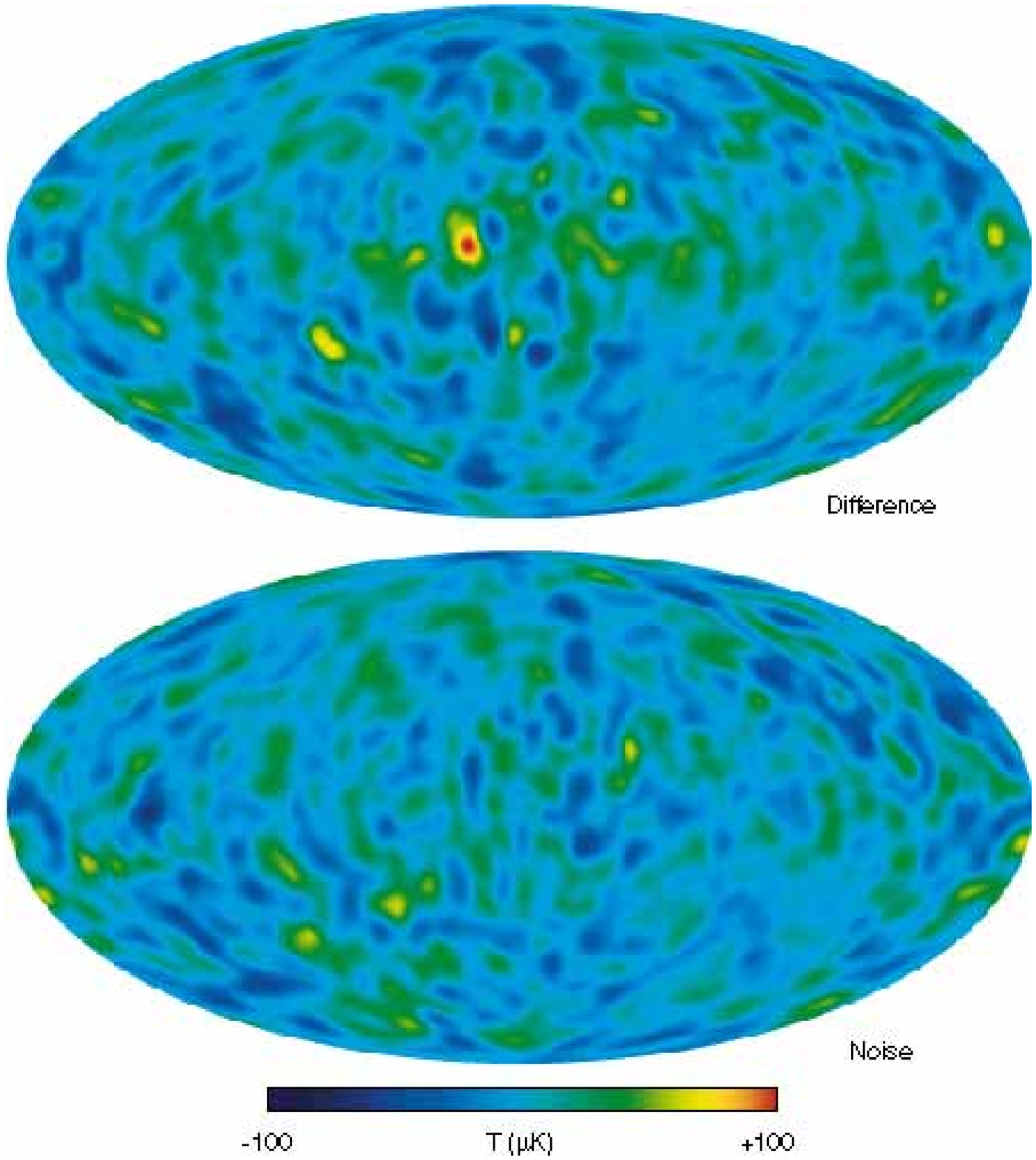}
\caption{The difference map is shown between the {\sl COBE}-DMR 53 GHz map and the combination 
Q-band/V-band maps from Figure {\ref{qv53}}.  This is compared with a map of the noise level.  
The maps are consistent with one another with the exception of a feature in the galatic
plane.  This discrepancy is likely to be due to a spectral index that is sufficiently different 
from the assumed CMB spectrum used to combine the \iMAP\ Q-band and V-band maps to mimic a 
53 GHz map.
\label{cobemapqv}} 
\end{figure}

Comparisons of data between \iMAP\ radiometers, and between \iMAP\ and {\sl COBE}, are 
important indicators of systematic error levels. 
Figure \ref{cobemapw} illustrates the enormous improvement in angular resolution 
from {\sl COBE} to \iMAP.  Features in the maps appear to be generally consistent, 
but the consistency is
better addressed by a more direct comparison.  To do this we take a combination of 
the \iMAP\ Q-band
and V-band maps and smooth it to mimic a {\sl COBE}-DMR 53 GHz map (see Figure \ref{qv53}).  
We then examine the difference between the {\sl COBE} and \iMAP\  pseudo-map.  Figure
\ref{cobemapqv} shows the difference map along with a simulated map of the noise.  
With the exception 
of a feature in the Galactic plane, the agreement is clearly at the noise level.  The Galactic 
plane feature is likely to be a result of the spectral 
index uncertainty of combining the Q-band and 
V-band maps to make a 53 GHz equivalent map.

\section{FOREGROUND ANALYSES \label{galaxy}}

\begin{figure}
\figurenum{10}
\epsscale{0.982}
\plotone{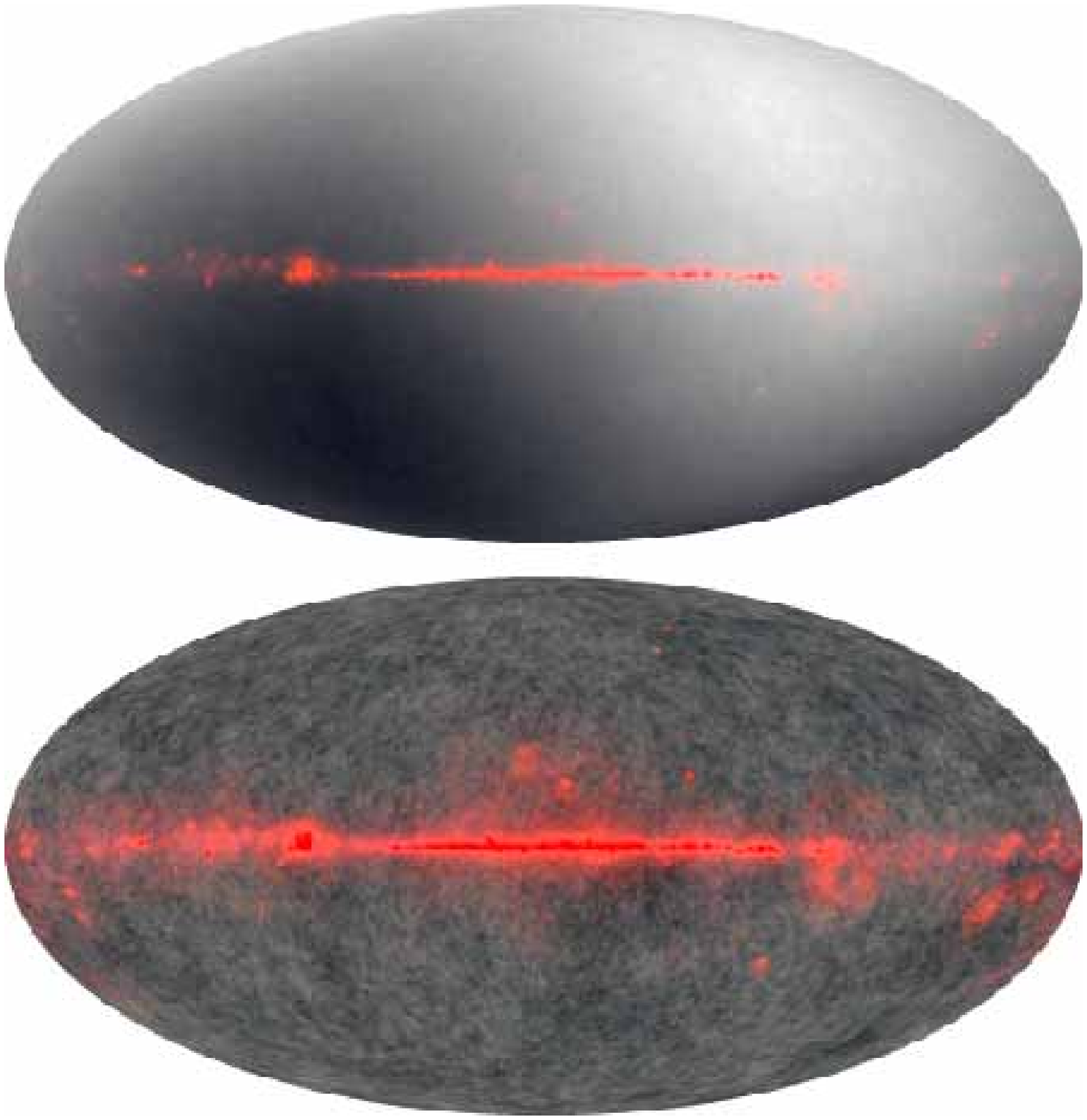}
\caption{False color images represent the spectral information from mutliple \iMAP\ bands.  
Q-band is red, V-band is green, and W-band is blue.  A CMB thermodynamic spectrum is grey.
({\sl top}) A three color combination image from the Q-, V-, and W-band maps.  The dipole and high
Galactic latitude anisotropy are seen.  
({\sl bottom}) A similar false color image but with the dipole subtracted.
\label{3band}} 
\end{figure}

An understanding of diffuse Galactic emission and extragalactic point sources is necessary 
for CMB analyses.  The \iMAP\ mission carries radiometers at five frequencies for the 
purpose of separating the CMB anisotropy from foreground emission based on their different 
spectra.  Figure \ref{3band} illustrates the spectral difference between the CMB and 
foregrounds.  The \iMAP\ bands were selected to be near the frequency where the ratio of the CMB
anisotropy to the contaminating foreground is at a maximum.

\subsection{Masks}

For CMB analyses it is necessary to mask out regions of bright foreground emission. 
\citet{bennett/etal:2003c} present a recipe for foreground masks based on 
K-band temperature levels.  Since foreground contamination is most severe in K-band, it  
is used as the best tracer of contamination.  The contamination morphology is similar enough 
across all five \iMAP\ bands that masks based on the temperature levels in other bands would be
redundant and unnecessary.  Standard names are given for the mask levels.  For example, 
the Kp0 mask cuts 21.4\% of sky pixels while the Kp2 mask cuts 13.1\%.  See
\citet{bennett/etal:2003c} for further detail.  An extragalactic point source mask is 
also constructed based on selections from source catalogs.  An additional 2\% of pixels 
are masked due to these $\sim 700$ sources.

\subsection{Diffuse Galactic emission}

\begin{figure}
\figurenum{11}
\epsscale{0.68}
\plotone{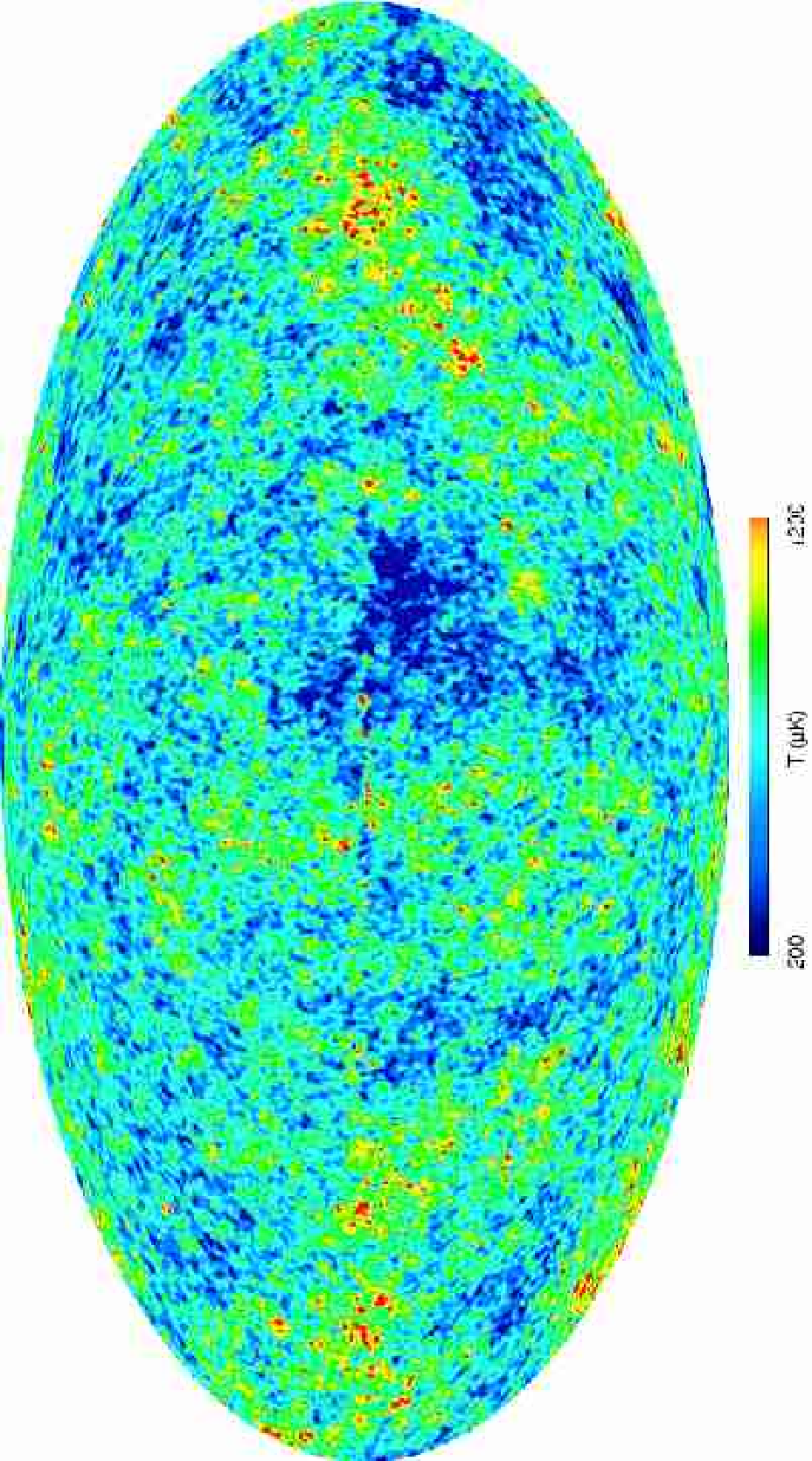}
\caption{This ``internal linear combination'' map combines the five band maps in such 
as way as to maintain unity response to the CMB while minimizing foreground foreground 
contamination.  For a more detailed description see \citet{bennett/etal:2003c}.  
For the region that covers the full sky outside of the inner Galactic plane, 
the weights are $0.109,   -0.684,   -0.096,   1.921,   -0.250$ for K, Ka, Q, V, and W bands, 
respectively.  Note that there is a chance alignment of a particularly warm feature and a 
cool feature near the Galactic plane.  As discussed in \citet{bennett/etal:2003c}, the 
noise properties of this map are complex, so it should not generally be used for cosmological 
analyses.
\label{ilc}} 
\end{figure}

Beyond the use of masks, one technique for reducing the level of foreground contamination 
is to form a linear combination of the multifrequency \iMAP\ data that retains unity response 
for only the emission component with a CMB spectrum.  This technique was introduced  
for {\sl COBE} by \citet{bennett/etal:1992b}.  With five \iMAP\ bands instead of the three on 
{\sl COBE}, and with a somewhat more elaborate approach for {\sl WMAP}, \citet{bennett/etal:2003c} 
arrive at the internal (\iMAP\ data only) linear combination map seen in Figure \ref{ilc} 
of this paper.  The foregrounds are removed to a remarkable degree; however, the statistics 
of this internal linear combination map are complex and inappropriate for most CMB analyses.

Below, we use the notation convention that flux density is 
$S\sim \nu^\alpha$ and antenna temperature is $T_A\sim \nu^\beta$, where the spectral 
indices are related by $\beta=\alpha-2$.  
In general, the CMB is expressed in terms of thermodynamic temperature, while Galactic and
extragalactic foregrounds are expressed in antenna temperature.  Thermodynamic temperature 
differences are given by $\Delta T = \Delta T_A [(e^x-1)^2/x^2e^x]$, where $x=h\nu/kT_0$, 
$h$ is the Planck constant, $\nu$ is frequency, $k$ is the Boltzmann constant, and 
$T_0=2.725$ K is the CMB temperature \citep{mather/etal:1999}.  Values of 
$\Delta T / \Delta T_A$ for the \iMAP\ bands are given by \citet{jarosik/etal:2003b} and 
can be found in Table \ref{overview}.

\citet{bennett/etal:2003c} identify the amplitudes and spectral indices of the 
individual emission components.  A maximum entropy method (MEM) approach is adopted 
where priors are used for component amplitudes and spectral indices, except for 
free-free emission, which has a fixed spectral index ($\beta=-2.15$ in the \iMAP\ bands).  
An iterative 
fit is performed, where the pixel-by-pixel amplitudes are updated in accordance with the MEM 
residuals until low ($<1$\%) residuals are achieved.  The process results in a map of each 
emission component for each of the five \iMAP\ bands.  The derived maps of thermal emission from
dust give a uniform spectral index across the sky of $\beta_d\approx 2.2$.  
The derived map of free-free 
emission is reasonable given the amplitude and morphology of H$\alpha$ measurements.  The other 
radio component fit should include 
the combined emission of synchrotron and spinning dust.  It shows the synchrotron 
spectrum steepening with increasing frequency, as would be expected for a spectral break 
due to synchrotron losses at $\approx 20$ GHz.  There is no indication of the less steep 
or flattening spectral index that would result from spinning dust emission.  The spinning 
dust emission is limited to $<5$\% of the total Ka-band foreground.  Reports of dust-correlated 
microwave emission from {\sl COBE} data analyses 
are understood as an admixture of the fraction of 
synchrotron emission (with $\beta\approx -3$) that is traced by a dust template, and thermal 
dust emission ($\beta\approx 2.2$), giving a combined spectral index of $\beta \approx -2.2$
between the \iMAP\ Ka-band and V-band, approximating the {\sl COBE} 31 GHz and 53 GHz bands.

While the MEM method is useful for understanding the nature of the foreground emission
components, these results can not be directly used in CMB analyses due to the complex noise 
properties that result from the MEM process and its simultaneous use of multifrequency maps.
This is because the multifrequency maps are smoothed, different weights are used in different 
regions of the sky and these weights are smoothed,  all of which complicates the noise 
correlations.
For the CMB analyses we use a mask to exclude pixels where the Galactic 
emission is strong, combined with template fitting (using external data only) where the 
foregrounds can be adequately corrected. This approach does not complicate the noise 
properties of the maps.  The Kp2 cut is used for all analyses except for limits on 
non-Gaussianity and the temperature-polarization correlation function, where the more 
severe Kp0 cut is used.

\citet{bennett/etal:2003c} describe the template fitting in detail.  
Thermal dust emission has been mapped over the full sky in several infrared bands, 
most notably by the {\it COBE} and {\it IRAS} missions.  A full sky template 
is provided by \citet{schlegel/finkbeiner/davis:1998}, 
and is extrapolated in frequency by 
\citet{finkbeiner/davis/schlegel:1999}.
The mostly synchrotron emission map of \cite{haslam/etal:1981} at 408 MHz is used as a radio
template.  The free-free ionized gas is traced by the H$\alpha$ map assembled by 
\citet{finkbeiner:2003} from the Wisconsin H-Alpha Mapper (WHAM), the Virginia Tech 
Spectral-Line Survey (VTSS), and the Southern H-Alpha Sky Survey Atlas (SHASSA) 
\citep{dennison/simonetti/topasna:1998, haffner/etal:2002, reynolds/haffner/madsen:2002, 
gaustad/etal:2001}.  The Haslam map resolution is not as high as that of the WMAP maps, 
and the Haslam map has artifacts from experimental effects such as striping from spatial 
calibration variations.  The striping in the Haslam map is along the survey scan lines 
and was corrected to first order by the application of a Wiener filter.  (The filtered 
version of the Haslam map is publicly available on the LAMBDA web site.)  The remaining 
adverse effects of the Haslam map are mitigated by two effects.  First, the template 
fit calls only for a small Haslam correlation (see \S6 of \citet{bennett/etal:2003c}).  
Since the correction is small, the error on the correction is negligible.  Second, the 
foreground contamination is most significant only on the largest angular scales so
the Haslam resolution limit and small-scale map artifacts are not significant sources 
of error.  The MEM solution only uses the Haslam map as a prior and the spinning dust 
limit only uses the full sky median of the Haslam map.  Thus the spinning dust limit 
is insensitive to residual striping in the Haslam map.

The MEM results are used to assess the degree of foreground emission remaining after the 
template subtraction.  The result is $<7\;\mu$K rms at Q-band and $<3\;\mu$K rms at both 
V-band and at W-band for $l<15$.  This remaining foreground emission constitutes $<2$\% 
of the CMB variance (up to $l=200$) in Q-band, and $<\atop\sim$1\% of the CMB variance 
in V- and W-bands.  Figures 3 and 4 of \citet{hinshaw/etal:2003} demonstrate that this 
small residual foreground level has a negligible effect on the cosmological results.

\subsection{Point sources}

A search was made for point sources in the \iMAP\ data. A catalog of 208 detected sources 
(with 98\% reliability) is provided by \citet{bennett/etal:2003c}.  Statistically, five 
sources are expected to be false detections.  Five of the 208 sources do not have low 
frequency radio counterparts; these sources are likely to be the false detections.
We include $\sim$700 sources in our mask, despite having only detected $\sim$200 sources 
at the $5\sigma$ level, 
because sources below this detection level still contribute an undesirable 
statistical contamination to the maps.  Even beyond masking the ~700 sources, we still 
need to make a statistical correction to the power spectrum for residual source contamination 
\citep{hinshaw/etal:2003}.
The derived source counts give a power spectrum level of 
$C^{\rm src} = (15\pm 3)\times 10^{-3}\; \mu$K$^2$sr at Q-band.  This is consistent with 
the level found in the bispectrum analysis of the maps \citep{komatsu/etal:2003} 
and the level found in fits to the map power spectra \citep{hinshaw/etal:2003}.
We have confidence that the point source level is understood since it is independently 
derived using three different methods.

\subsection{Sunyaev-Zeldovich (SZ) effect}

Hot gas in clusters of galaxies imparts energy to the CMB photons, causing a temperature 
decrement in the \iMAP\ bands (the Sunyaev-Zeldovich Effect).  The Coma cluster is expected 
to have the most pronounced signature.  For the highest resolution maps, 
\cite{bennett/etal:2003c} get 
$-0.34\pm 0.18$ mK in W-band and 
$-0.24\pm 0.18$ mK in V-band in the direction of Coma.  
Use of the XBACS catalog of X-ray clusters as a template results in $-0.36\pm 0.14$.  
This verifies that the Sunyaev-Zeldovich Effect is barely detectable in even a matched 
search of the \iMAP\ sky maps and therefore it is not a significant ``contaminant'' 
to the \iMAP\ data.

\section{LIMITS ON NON-GAUSSIANITY\label{gauss}}

Maps of the sky are the most complete and compact representation of the CMB anisotropy.
Cosmological analyses are based on statistical properties of the maps with the power 
spectrum as one of the most commonly derived statistics.  
The power spectrum is a complete representation of the data only if the CMB anisotropy 
is Gaussian.  
Also, the most common cosmological models predict that the CMB anisotropy should be 
consistent with a Gaussian random field 
(at least at levels that are currently possible to measure).  Therefore, 
we test the Gaussianity of the anisotropy, both to interpret the power spectrum 
(and other statistical derivatives of the maps) and to test cosmological paradigms. 

There is no single best test for Gaussianity.  Specific tests can be more or less sensitive to
different assumed forms of non-Gaussianity.
\citet{komatsu/etal:2003}, in a companion paper, search for non-Gaussianity in 
the \iMAP\ CMB anisotropy maps using Minkowski functionals and a bispectrum estimator.  

Minkowski functionals \citep{minkowski:1903, gott/etal:1990} quantify topological 
aspects of the CMB maps.  Anisotropy is examined via contours at different 
temperature levels, and the number and areas of regions enclosed by these contours 
are computed.  Three Minkowski functionals are the area represented by hot and cold spots, 
the contour length around these areas, and the difference between the number of these 
areas (the ``genus'').

It is widely believed that the CMB anisotropy arises from Gaussian linear fluctuations in 
the gravitational potential.  \citet{komatsu/spergel:2001} suggest that non-Gaussian 
anisotropy be considered in terms of the curvature perturbation.  The simplest 
expression for the overall primordial gravitational curvature perturbation, 
$\Phi({\bf x})$, is a sum of a linear $\Phi_L({\bf x})$ and weak nonlinear components:
$\Phi({\bf x}) = \Phi_L({\bf x}) + f_{NL} [\Phi_L^2({\bf x}) - \left< \Phi_L({\bf x})\right>^2]$
where $\Phi_L$ is the linear Gaussian portion of the curvature perturbation 
and $f_{NL}$ is a nonlinear coupling constant.  Then, $f_{NL}=0$ 
corresponds to the purely linear Gaussian case.
Since the CMB bispectrum measures the phase 
correlations of the anisotropy, it can be used to solve for $f_{NL}$.  The Minkowski 
functional results can also be expressed in terms of $f_{NL}$.

For the Minkowski functionals, \citet{komatsu/etal:2003} find 
$f_{NL} < 139$ (95\% CL).  From the bispectrum, \citet{komatsu/etal:2003}
find $-58<f_{NL}<134$ (95\% CL).  The two results are consistent.
The CMB anisotropy is thus demonstrated to follow Gaussian statistics.
This is a significant result for models of the early universe.
It also means that we can construct and interpret CMB statistics 
(e.g. the angular power spectrum) from the maps in a straightforward manner.  

\section{MULTIPOLES \label{multipoles}}

The temperature anisotropy, $T({\bf n})$, is naturally 
expanded in a spherical harmonic 
basis, $Y_{lm}$, as 
\begin{equation}
T({\bf n}) = \sum\limits_{l,m} a_{lm} Y_{lm}({\bf n}).
\end{equation}
The angular power spectrum, $C_l$, is a cosmological ensemble average given by
\begin{equation}
C_l=\langle \vert a_{lm}\vert^2 \rangle
\end{equation}
and is observed for our actual sky as
\begin{equation}
C_l^{sky}={\frac{1}{2l+1}}\sum \limits_m \vert a_{lm} \vert ^2.
\end{equation}
Assuming random phases, the temperature anisotropy for each multipole moment, $\Delta T_l$, 
can be associated with the angular spectrum, $C_l$, as
\begin{equation}
\Delta T_l = \sqrt{C_l^{sky}\; l(l+1) / 2\pi}.
\end{equation}
The correlation function is
\begin{eqnarray}
C(\theta) & = & {\frac{1}{4\pi}} \sum\limits_l \sum\limits_{m=-l}^{+l} C_l P_l(cos\; \theta) \\
          & = & {\frac{1}{4\pi}} \sum\limits_l (2l+1) C_l\; P_l(cos\; \theta)
\end{eqnarray}
where $P_l$ is the Legendre polynomial or order $l$.

In practice, an instrument adds noise and spatially filters the sky signal due to the beam pattern
and any other experimental limitations on the sampling of all angular scales. 
The experimental transfer function, $b_l$, for each multipole moment depends upon the specific 
beam pattern of the experiment.  \iMAP\ samples all angular scales limited only by the beam 
pattern.  The window function for the signal power between channels $i$ and $j$ is 
\begin{equation}
w_l^{ij} = b_l^i b_l^j p_l^2,
\end{equation}
where $p_l^2$ is the pixel window function.
The angular power spectrum then becomes
\begin{equation}
C_l = C_l^{ij} b_l^i b_l^j p_l^2 + N_l^i\delta_{ij}  \label{ps-eqn}
\end{equation}
where $N_l^i$ is the power spectrum noise that results from the instrument noise, 
which is assumed to be uncorrelated between channels $i$ and $j$.

Note that an auto-power spectrum has twice the noise variance as a cross-power 
spectrum with the corresponding noise in each map. This follows from the 
property of Gaussian noise that 
$\left< N_l^{i~2} \; N_l^{j~2}\right> =  2 \left< N_l^{i~2} \right>^2 \delta_{ij} 
   + \left< N_l^{i~2} \right> \left< N_l^{j~2} \right>$, with 
$\left<N_l^i N_l^j\right>^2 = \left<N_l^{i~2}\right>^2 \delta_{ij}$, 
and 
$\left< N_l^{i~4} \right> = 3 \left< N_l^{i~2} \right> ^2$.

The use of a sky mask for foreground suppression breaks the orthogonality of the 
spherical harmonics on the sky and leads to mode coupling.  \citet{hinshaw/etal:2003}
discuss how \iMAP\ handles this, and other complexities. 

\subsection{$l=1$ dipole \label{dipole}}

{\sl COBE} determined the dipole amplitude is $3.353\pm 0.024$ mK in the direction 
$(l,b)=(264.26^\circ\pm0.33^\circ, 48.22^\circ\pm 0.13^\circ)$, where $l$ is Galactic longitude 
and $b$ is Galactic latitude \citep{bennett/etal:1996}. 
This dipole was subtracted from the \iMAP\ data during processing. 
Examination of the \iMAP\ maps allow for the determination of a residual dipole, and thus an
improvement over the {\sl COBE} value.  Note that this does not have any effect on \iMAP\ 
calibration, which is based on the Earth's velocity modulation of the dipole, and not
on the dipole itself.  The \iMAP\ -determined dipole is $3.346\pm 0.017$ mK in the direction 
($l$, $b$)$=$ ($263\ddeg 85\pm0 \ddeg 1$, $48\ddeg 25\pm0\ddeg 04$).  The uncertainty of the 
dipole amplitude is limited by the \iMAP\ 0.5\%
calibration uncertainty, which will improve with time.

\subsection{$l=2$ quadrupole \label{quadrupole}}

The quadrupole is the $l=2$ term of the spectrum $\Delta T_l^2 = l(l+1)C_l/2\pi$, i.e. 
$\Delta T_{l=2}^2=(3/\pi)C_{l=2}$.  Alternately, the quadrupole amplitude can be expressed as
$Q_{rms}=\sqrt{(5/4\pi)\;C_{l=2}} = \sqrt{5/12}\;\Delta T_{l=2}$.
The 4-year {\sl COBE} quadrupole is $Q_{rms}=10^{+7}_{-4}\;\mu$K with the peak of the 
likelihood in the range $6.9\;\mu{\rm K} < Q_{rms} < 10\;\mu$K, 
as shown in Figure 1 of \citet{hinshaw/etal:1996a}.  

The \iMAP\ quadrupole, $Q_{rms}=8\pm 2\; \mu$K or $\Delta T_2^2=154\pm70$ $\mu$K$^2$, is 
consistent with {\sl COBE} but with tighter limits because of better measurements and
understanding of foregrounds.  
We determine the quadrupole value by computing the power spectrum 
of the internal linear combination map and individual channel maps, with 
and without foreground corrections, for a range of Galactic cuts.  
The final $l=2$ value corresponds to a full sky estimate with an 
uncertainty that encompasses a range of foreground-masked or 
foreground-corrected solutions."
The foreground level is still the leading uncertainty.  
(The small kinematic quadrupole 
is not removed from the maps nor accounted for in this analysis.)  
The quadrupole value is low compared with values predicted 
by $\Lambda$CDM models that fit the rest of the power spectrum.  $\Lambda$CDM models, in
particular, tend to predict relatively high quadrupole values due to the enhanced, 
$\Lambda$-driven, integrated Sachs-Wolfe effect.

\subsection{n-poles \label{npoles}}

A central part of the task of computing multipole information from the maps 
is the evaluation 
and propagation of errors and uncertainties.  This largely involves arriving at 
an adequate representation of the Fisher matrix, which is the inverse covariance 
matrix of the data.  The Fisher matrix must take into account mode-coupling 
from the sky cut, beam (window function) uncertainties, and noise properties.
The fact that the \iMAP\ data use a nearly azimuthally symmetric cut, and have 
a nearly diagonal pixel-pixel covariance, greatly simplifies the evaluation 
of the Fisher matrix.  

Two approaches to computing the angular power spectrum 
have been used by \cite{hinshaw/etal:2003}: a quadratic estimation 
based on \citet{hivon/etal:2002}; and a maximum likelihood estimate based on 
\citet{oh/spergel/hinshaw:1999}.  The 
quadratic estimator is used in the final \iMAP\ spectrum analyses,
 while the maximum likelihood technique is used as a cross-check.

The K-band and Ka-band beam sizes are large enough that these bands are not 
used for CMB analysis since they have the most foreground contamination and 
probe the region in $l$-space that is 
cosmic variance limited by the measurements at the other bands.  
These bands are invaluable, however, as monitors of 
Galactic emission.  The two Q-band, two V-band, and four W-band differencing 
assemblies are the source of the prime CMB data.  The matrix of auto- and 
cross-correlations between the eight Q-, V-, and W-band differencing assemblies 
has eight diagonal (auto-correlations) and 28 unique off-diagonal elements
(cross-correlations). 
Since auto-correlations are difficult to assess due to the noise bias (see equation 
\ref{ps-eqn}), 
we included only the 28 unique off-diagonal (cross-correlations) in the \iMAP\ 
power spectrum analysis.  In dropping the auto-correlations, each of which has twice the 
noise variance of a cross-correlation, we lose only $1-\sqrt{56/(56+8)}=6$\% 
of the ideally achievable signal-to-noise ratio.  In \citet{hinshaw/etal:2003}, 
we show that the power spectrum computed from the auto-correlation data is 
consistent with the angular power spectrum from the cross-correlation data.  We anticipate 
using the auto-correlation data in future analyses.

The cross-power spectrum from the 28 pairs is considered in four $l$ ranges. 
For $l\le 100$ we use uniform pixel weighting of only V- and W-band data. This 
reduces the Galactic contamination where measurement errors are well below the 
cosmic variance. For $100<l\le 200$ we use uniform pixel weighting of the combined 
28 pairs.  
For $200<l\le 450$ all 28 cross-power pairs are used with 
a transitional pixel weighting. The transitional pixel weighting, defined and 
discussed in detail in Appendix A of \citet{hinshaw/etal:2003}, smoothly 
transitions the weighting from the uniform pixel weights in the signal-dominated 
$l<200$ multipole regime to inverse-noise-variance weighting in the noise-dominated 
$l>450$ multipole regime.  For $l>450$ all 28 pairs are used with 
inverse noise weighting. Our Monte Carlo simulations show that this 
approach is a nearly optimal scheme.

\begin{figure}
\figurenum{12}
\epsscale{.7}
\plotone{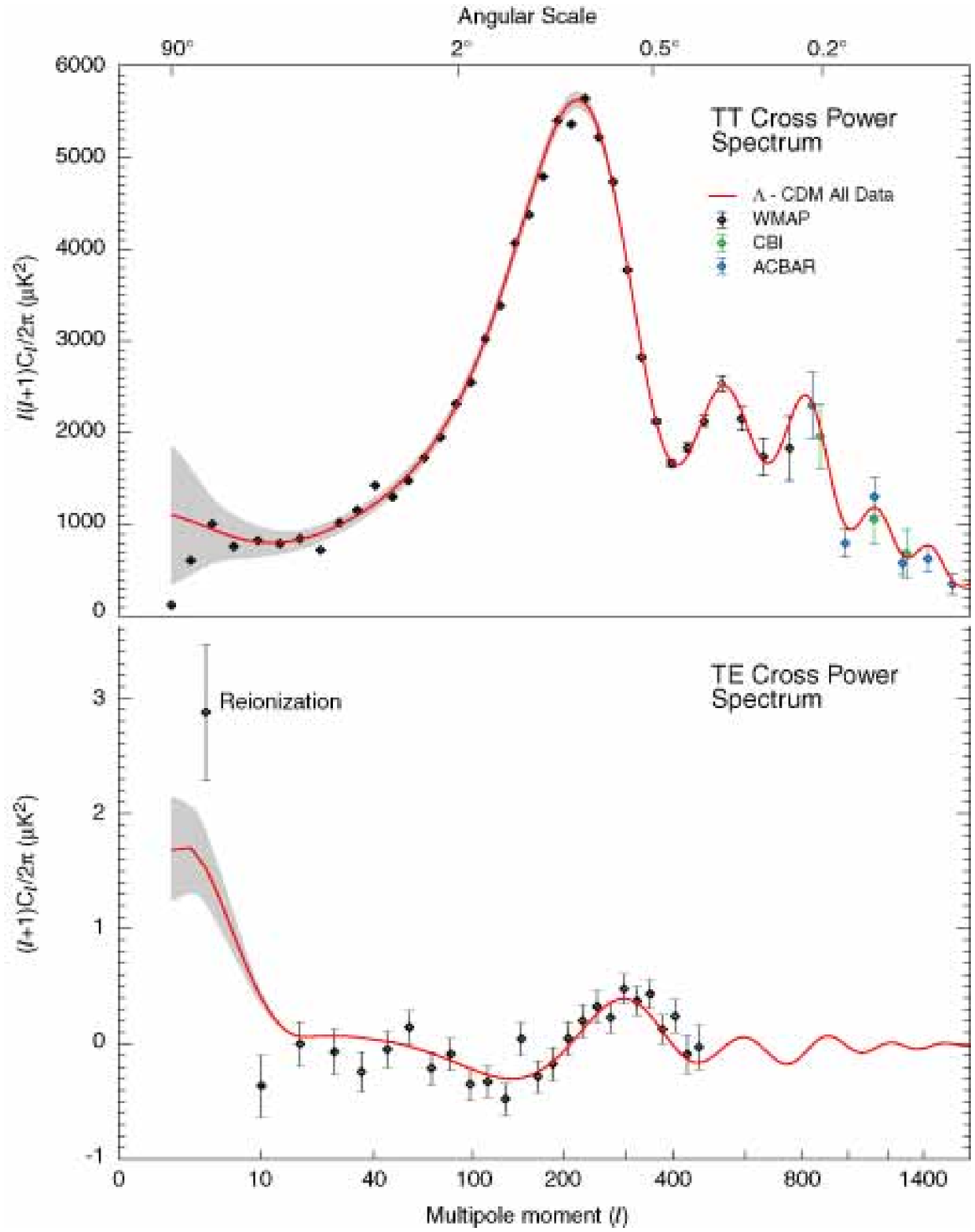}
\caption{The \iMAP\ angular power spectrum.  ({\sl top}:)  
The \iMAP\ temperature (TT) results are consistent with the ACBAR and CBI measurements, 
as shown.  The TT angular power spectrum is now highly constrained.  
Our best fit running index $\Lambda$CDM model is shown.  The grey band
represents the cosmic variance expected for that model.  
The quadrupole has a surprisingly low amplitude.  
Also, there are excursions from a smooth spectrum (e.g., at 
$\ell\approx 40$ and $\ell\approx 210$) 
that are only slightly 
larger than expected statistically. While intriguing,
they may result from a combination of cosmic variance, subdominant
astrophysical processes, and small effects from approximations
made for this first year data analysis \citep{hinshaw/etal:2003}.
We do not attach cosmological significance to them at present.  
More integration time and more detailed analyses are needed.
({\sl bottom}:) 
The temperature-polarization (TE) cross-power spectrum, $(l+1)C_l/2\pi$.  
(Note that this is {\it not} multiplied by the additional factor of $l$.)
The peak in the TE spectrum near $l\sim 300$ is out of phase with the TT power
spectrum, as predicted for adiabatic initial conditions.  The antipeak in the 
TE spectrum near $l\sim 150$ is evidence for superhorizon modes at decoupling, 
as predicted by inflationary models.
\label{ps}} 
\end{figure}

The angular power spectrum is shown for the \iMAP\ data in Figure \ref{ps}.  The 
\iMAP\ power spectrum agrees closely with {\sl COBE} at the largest angular scales, and with 
CBI and ACBAR at the finer angular scales.  We highlight the CBI and ACBAR results 
because they are a useful complement 
to \iMAP\ at the smaller angular scales.  The acoustic pattern is obvious.  
\citet{page/etal:2003c} find that the first acoustic peak is 
$\Delta T_l=74.7\pm 0.5\; \mu$K at $l=220.1\pm 0.8$.  
The trough following this peak is 
$41.0\pm 0.5\; \mu$K at $l=411.7\pm 3.5$ and the second peak is 
$48.8\pm 0.9\; \mu$K at $l=546\pm 10$.  

\begin{figure}
\figurenum{13}
\epsscale{.5}
\plotone{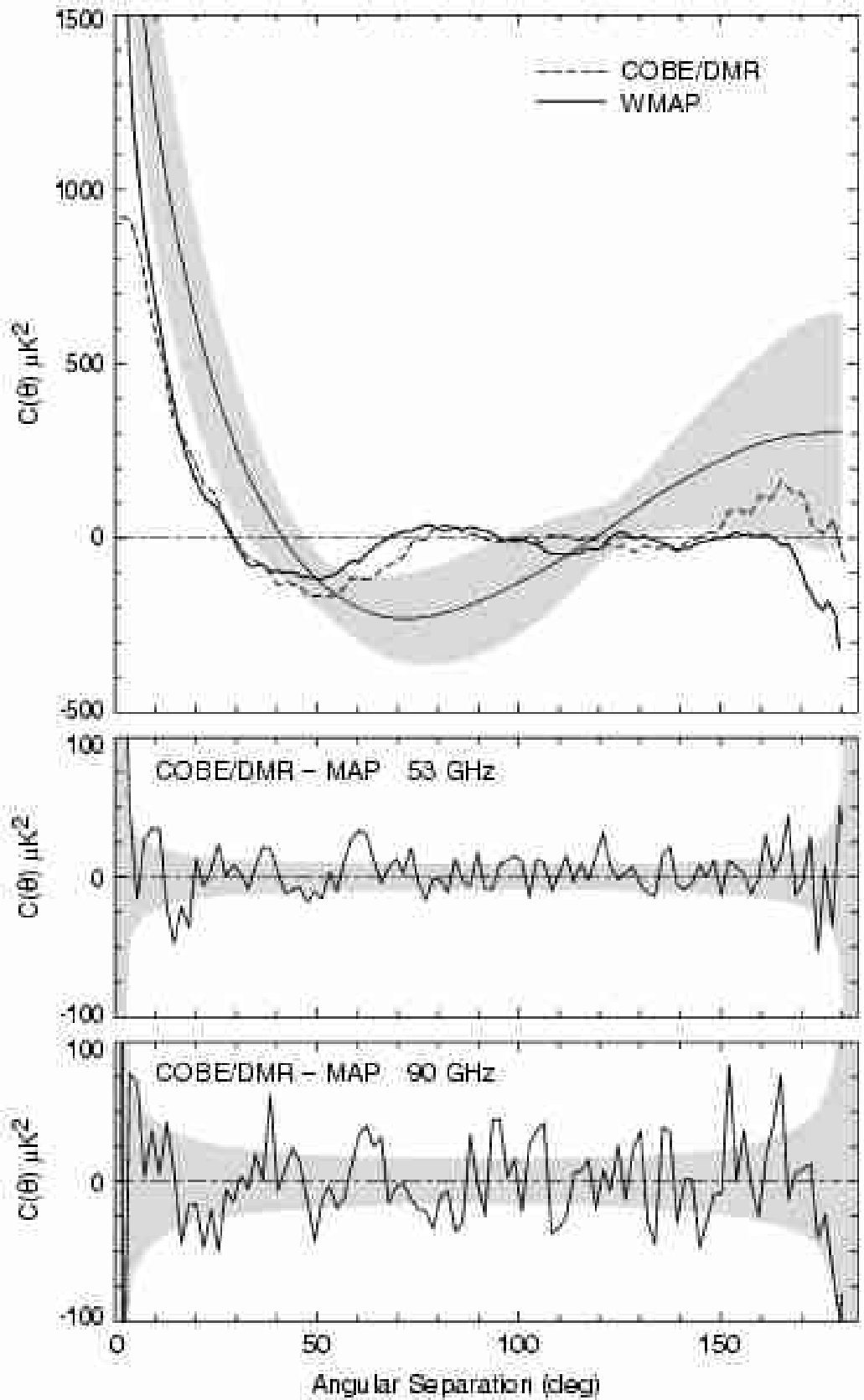}
\caption{{\it (top): }CMB temperature correlation function 
of the \iMAP\ and {\sl COBE} data.  
The \iMAP\ correlation function is computed using a combination of the 
Q-band, V-band, and W-band maps with the Kp0 cut sky and the MEM Galactic model 
subtracted. The {\sl COBE} correlation 
function is computed using the ``custom cut'' sky.  The running index $\Lambda$CDM model 
that is fit to the power spectrum 
is shown with a Monte Carlo determined grey band indicating one standard deviation.  
The model is, overall, an excellent fit to the \iMAP\ power spectrum.  However, a 
correlation plot emphasizes the low $l$ power.
The discrepancy between the model and data illustrates that there is surprisingly little 
anisotropy power in the \iMAP\ and {\sl COBE} maps at large angles.  
{\it (bottom): } The lower two plots display the correlation function of the difference between 
the {\sl COBE}-DMR and \iMAP\ maps with a $\vert b\vert = 10^\circ$ Galactic plane cut.  
A synthesis of the \iMAP\ Q and V band maps was made to
approximate a 53 GHz-like map to compare with the {\sl COBE}-DMR 53 GHz map.  The {\sl COBE}-DMR 
90 GHz map is compared directly, without corrections, to the \iMAP\ W-band map.  
These plot emphasize the consistency of the \iMAP\ and {\sl COBE} measurements.  
The slightly higher than expected deviations at 53 GHz are likely to be due to Galactic
contamination, arising from outside the cut regions and from the construction of the synthesized 
\iMAP\ 53 GHz map.
\label{corr}} 
\end{figure}

$\Lambda$CDM models predict enhanced large angle power due to the integrated Sachs-Wolfe effect.  
The \iMAP\ and {\sl COBE} data, on the other hand, have the opposite trend.
The conflict is also seen clearly in the correlation function, $C(\theta)$, shown  
in Figure \ref{corr}.  The \iMAP\ correlation function is computed using the Kp0 cut on 
a combination of the Q-band, V-band, and W-band maps with the MEM Galactic model removed. 
The {\sl COBE} correlation function is computed on the ``custom cut'' sky 
\citep{bennett/etal:1996}.  The best-fit 
$\Lambda$CDM model is shown with a grey band indicating one standard 
deviation as determined by Monte Carlo simulations.  

The lower two plots in Figure \ref{corr} display the correlation function of the difference 
between the {\sl COBE}-DMR and \iMAP\ maps with a $\vert b\vert = 10^\circ$ Galactic plane cut.  
A synthesis of the \iMAP\ Q- and V-band maps was made to
approximate a 53 GHz map to compare with the {\sl COBE}-DMR 53 GHz map.  The {\sl COBE}-DMR 
90 GHz map is compared directly, without corrections, to the \iMAP\ W-band map.  
These plots emphasize the consistency of the \iMAP\ and {\sl COBE} measurements.  
The slightly higher than expected deviations at 53 GHz are likely to be due to Galactic
contamination, arising from outside the cut regions and from the construction of the synthesized 
\iMAP\ 53 GHz map.

The model is an excellent fit to the \iMAP\ full power spectrum except, perhaps, 
at $l \lsim 6$.  Since only a small fraction of the total number of measured 
multipoles are involved, the statistical contribution of the $l \lsim 6$ points to the 
overall power spectrum fit is small and does not greatly drive the overall best-fit model.
The correlation function emphasizes the low $l$ signal because these modes contribute 
to $C(\theta)$ at all angular separations.  The discrepancy at 
$\theta \lsim 30^{\circ}$ reflects the average lack of power in the 
data relative to the model at $l<6$.  More significantly, the lack 
of power at $\theta \gsim 60^{\circ}$ relative to the model reflects the special 
shape of the power spectrum from $2<l<5$ seen in both the \iMAP\  (and {\sl COBE}) 
maps.  This result is generically 
true for $\Lambda$CDM models, independent of the exact parameters of the model.
{\it There is very little large scale CMB anisotropy power in our sky.  
This fact, first seen by 
{\sl COBE} is confirmed by \iMAP.}  The probability of so little $C(\theta > 60^\circ)$ 
anisotropy power is $\sim 2\times 10^{-3}$, given the best-fit $\Lambda$CDM model.  
The lack of power 
is seen both in $C(\theta)$ and the behavior of the low order 
($l=2, 3, 4, {\rm and}\; 5$) multipoles.

\section{CMB POLARIZATION \& THE DETECTION OF REIONIZATION \label{reion}}

Each differencing assembly measures the sky in two orthogonal linear polarizations.
As the Observatory spins,  precesses, and orbits the Sun, the instrument observes 
the sky over a range of polarization angles.  The range of angles observed is 
neither complete nor uniform, but it is sufficient to provide valuable new CMB 
polarization results.

We express polarization in term of the standard $I$, $Q$, and $U$ Stokes parameters, 
in units of thermodynamic temperature.
Polarization coordinates are not rotationally invariant so a coordinate system is 
defined by \citet{kogut/etal:2003} for expressing information.
By appropriate summing 
and differencing of the time-ordered-data from the pairs of corresponding radiometers,  
we iteratively make maps in unpolarized intensity, $I$, and in each polarized 
component, $Q$ and $U$ \citep{hinshaw/etal:2003}.  
The $Q$ and $U$ maps not only have much lower signal-to-noise
than the $I$ maps, but they are more susceptible to systematic errors 
since the differencing between radiometer outputs occurs on the ground, not in the 
front-end of the radiometers.

The $U$ polarization maps are expected to be most affected by systematic errors. 
This is due to the orientation of the observed polarization angles relative to the spin 
axis of the Observatory, causing systematics not fixed on the sky to preferentially 
go into the $U$ map.  

Taurus A, an extremely strong source in the \iMAP\ bands, is polarized and its polarization 
amplitude and direction are known \citep{flett/henderson:1979}.  
Polarization measurements from \iMAP\ are consistent 
with previous observations of Taurus A, providing a useful check on the operation 
of the \iMAP\ hardware and software.
However, the polarization systematic errors have not yet been fully quantified.
We chose to release the $I$ data and results as soon as possible,
rather than postpone their publication and release until the systematic measurement 
errors of the $Q$ and $U$ data are fully assessed.  All data, including the $Q$ and $U$ 
maps will be released when the characterization of their instrumental signatures 
are complete.  The dominant systematic effect in the low signal-to-noise 
polarization data arises from correlated noise in the radiometers.  By using only 
cross correlations of temperature maps with polarization maps, generated from 
independent radiometer combinations that have uncorrelated noise, we mitigate 
the leading systematic error.
For this reason, results of temperature-polarization correlations between the $I$ and $Q$ 
maps, i.e. TE correlations,
are much less sensitive to systematic effects than polarization signals alone, 
so we are able to report those results.  

Figure \ref{ps} shows the temperature-polarization (TE) cross power spectrum.  
\citet{kogut/etal:2003} report a detection of TE correlations on both large angular 
scales (from reionization) and on small scales (from the adiabatic fluctuations). 
The TE power spectrum, shown in Figure \ref{ps}, is discussed in detail 
by \citet{kogut/etal:2003}.  In the TE angular power spectrum the 
antipeak is $-35\pm 9\;\mu$K$^2$ at $l=137\pm 9$ and the peak is 
$105\pm 18\;\mu$K$^2$ at $l=329\pm 19$ \citep{page/etal:2003c}.

The detection of the reionization of the universe corresponds to an optical depth  
\ensuremath{\tau = 0.17 \pm 0.04} 
($0.09\le \tau \le 0.28$ at 95\% confidence).  
Although \iMAP\ measures the integrated optical depth, the epoch and redshift of 
reionization can 
be derived from the integral optical depth within the context of a model of the 
reionization process.  A single instantaneous step function in the ionization fraction
from zero to a steady 
fixed value is physically unlikely.  For more likely models, 
\citet{kogut/etal:2003} conclude that 
the redshift of reionization is 
$z_r=20^{+10}_{-9}$ (95\% CL), 
corresponding to an epoch of reionization of 
$t_r=180^{+220}_{-80}$ Myr (95\% CL) after the Big Bang.  \citet{cen:2002} 
presents a detailed model of a reionization process and predicts an integral 
value of the optical depth, $\tau=0.10\pm 0.03$.  This is low compared with the newly 
measured value, but would be higher and consistent if the assumed cosmological parameters 
in the model 
were adjusted to the values of the new \iMAP\ 
best-fit parameters (see \S\ref{model}).  

The measured optical depth means that reionization suppressed the acoustic peak 
amplitudes by $1-e^{-2\tau}\approx 30$\%.  While accounted for in our model fits 
(\S\ref{cosmo}, below), this suppression was not accounted for in previous CMB 
parameter determinations.  The anticorrelations observed in the TE power spectrum directly 
imply superhorizon fluctuations, a new result in support of inflation-like theories, 
as discussed in \S\ref{inflation}.

\section{COSMOLOGICAL INTERPRETATION\label{cosmo}}

In this section we summarize the cosmological intepretation of \iMAP\ 
first year results, which are discussed in more detail by \citet{spergel/etal:2003}, 
\citet{peiris/etal:2003}, \citet{page/etal:2003c}, and \citet{kogut/etal:2003}.  
The methodology used in the model fits is described by \citet{verde/etal:2003}.

\citet{spergel/etal:2003} show that a cosmological model with a flat universe,  
seeded with a scale-invariant spectrum of adiabatic Gaussian fluctuations, 
with reionization, is an acceptable fit not only to the \iMAP\ data but also to a 
host of astronomical data.  These data are:  
smaller angular scale CMB anisotropy data from ACBAR \citep{kuo/etal:2002} and CBI 
\citep{pearson/etal:2002};   
the HST key project value of $H_0$ \citep{freedman/etal:2001}; 
the accelerating Universe seen in Type Ia SNe \citep{riess/etal:2001};
the shape and amplitude of the large scale structure seen in clusters 
and superclusters of galaxies \citep{percival/etal:2001, verde/etal:2003}; 
and the linear matter power spectrum seen in the Lyman $\alpha$ forest 
\citep{croft/etal:2002}.
There has been mounting evidence in the direction of 
this model for years \citep{peebles:1984, bahcall/etal:1999}.
The optical depth since reionization  is a new, 
but not surprising component of the model.    
The \iMAP\ data establish this model as the standard model of 
cosmology by testing the key assumptions of the model and by enabling a precise 
determination of its parameters.

The \iMAP\ data test several of the key tenets of the standard model.
The \iMAP\ detection of temperature-polarization 
correlations \citep{kogut/etal:2003} and the clear
detection of acoustic peaks \citep{page/etal:2003c}
implies that the primordial fluctuations were primarily
adiabatic: the primordial ratio of dark matter/photons and the primordial ratio of
baryons/photons do not vary spatially.   The analysis of the \iMAP\ temperature data
demonstrates Gaussianity \citep{komatsu/etal:2003}.
The \iMAP\ data, when combined with any one
of the following three external data sets: 
HST Key Project measurement of $H_0$ \citep{freedman/etal:2001},
the 2dFGRS measurement of the matter density \citep{percival/etal:2001, verde/etal:2003} {\it or} 
the Type Ia supernova measurements \citep{riess/etal:2001}
imply that the radius of curvature of the universe, 
$R = c H_0^{-1} |1-\Omega_{tot}|^{-1/2}$, 
must be very large, \ensuremath{\Omega_{tot} = 1.02 \pm 0.02 \mbox{ }} and 
\ensuremath{0.99 < \Omega_{tot} < 1.05 \mbox{ }(95\%\mbox{\ CL})}.
These measurements
also require that the dark energy be the dominant constituent of 
$\Omega_{tot}$.  The \iMAP\ data alone
rule out the standard $\Omega_m = 1$ CDM model by $> 7\sigma$.

\subsection{Best Fit Cosmological Model \label{model}}

While an acceptable fit, the model described above is not our best fit model.  
In the discussion below we concentrate on our best fit model, which adds a 
scale-dependent primordial spectral index.  This cosmological model 
is a flat universe with a baryon
fraction of \ensuremath{\Omega_b = 0.044 \pm 0.004},
a matter fraction of \ensuremath{\Omega_m = 0.27 \pm 0.04},
and a dark energy fraction of \ensuremath{\Omega_\Lambda = 0.73 \pm 0.04},
seeded with a scale-dependent spectrum of adiabatic Gaussian fluctuations.  
This model has $C_{l=10}=46.0$ $\mu$K$^2$, consistent with the {\sl COBE} 
measurement of $C_{l=10}=44.4$ $\mu$K$^2$.

The \iMAP\ data alone enable accurate determinations of many of the key
cosmological parameters \citep{spergel/etal:2003}.  But a combination
of the \iMAP\ data with
the {\sl COBE} determination of the CMB temperature \citep{mather/etal:1999}, the CBI 
\citep{pearson/etal:2002} and the ACBAR \citep{kuo/etal:2002} CMB measurements,  
and the 2dFGRS survey determination
of the power spectrum of the local galaxy fluctuations \citep{percival/etal:2001}, 
yields the best fit cosmological
parameters listed in Table \ref{tbl-2}.  \citet{verde/etal:2003} describes our 
methodology for
determining these parameters and \citet{spergel/etal:2003} describes the best fit models
for different combinations of data sets.

\begin{deluxetable}{lcccc}
\tabletypesize{\footnotesize}
\tablecaption{``Best'' Cosmological Parameters  \label{tbl-2}}
\tablewidth{0pt}
\tablehead{
\colhead{Description} & \colhead{Symbol}   &  \colhead{Value}&  \colhead{$+$ uncertainty}&  \colhead{$-$ uncertainty}}
\startdata
Total density          &\ensuremath{\Omega_{tot}}
                       &\ensuremath{1.02} 
                       &\ensuremath{0.02}  
                       &\ensuremath{0.02}\\
Equation of state of quintessence &$w$ & $<-0.78$ & 95\% CL& ---\\
Dark energy density    &\ensuremath{\Omega_\Lambda}
                       &\ensuremath{0.73}
                       &\ensuremath{0.04}
                       &\ensuremath{0.04}\\
Baryon density         &\ensuremath{\Omega_bh^2} &\ensuremath{0.0224} 
                       &\ensuremath{0.0009}   &\ensuremath{0.0009}\\
Baryon density         &\ensuremath{\Omega_b} &\ensuremath{0.044} 
                       &\ensuremath{0.004}   &\ensuremath{0.004}\\
Baryon density (cm$^{-3}$) &\ensuremath{n_b} &\ensuremath{2.5 \times 10^{-7}} 
                       &\ensuremath{0.1 \times 10^{-7}}   &\ensuremath{0.1 \times 10^{-7}}\\
Matter density         &\ensuremath{\Omega_mh^2} &\ensuremath{0.135} 
                       &\ensuremath{0.008}   &\ensuremath{0.009}\\
Matter density         &\ensuremath{\Omega_m} &\ensuremath{0.27} 
                       &\ensuremath{0.04}   &\ensuremath{0.04}\\
Light neutrino density &$\Omega_\nu h^2$ & $<0.0076$ & 95\% CL & --- \\
CMB temperature (K)\tablenotemark{a}      &$T_{\rm cmb}$& 2.725 & 0.002 & 0.002 \\
CMB photon density (cm$^{-3}$)\tablenotemark{b}  &$n_\gamma$& 410.4 & 0.9& 0.9 \\
Baryon-to-photon ratio &\ensuremath{\eta} &\ensuremath{6.1 \times 10^{-10}} 
                       &\ensuremath{0.3 \times 10^{-10}}   &\ensuremath{0.2 \times 10^{-10}}\\
Baryon-to-matter ratio &$\Omega_b \Omega_m^{-1}$ & \ensuremath{0.17} 
                       &\ensuremath{0.01}   &\ensuremath{0.01}\\
Fluctuation amplitude in $8h^{-1}$ Mpc spheres    &\ensuremath{\sigma_8}     &\ensuremath{0.84}
                       &\ensuremath{0.04}       &\ensuremath{0.04}\\
Low-$z$ cluster abundance scaling &\ensuremath{\sigma_8\Omega_m^{0.5}}     &\ensuremath{0.44}
                       &\ensuremath{0.04}       &\ensuremath{0.05}\\
Power spectrum normalization (at $k_0=0.05$ Mpc$^{-1}$)\tablenotemark{c} 
                       &$A$& 0.833 & 0.086 & 0.083\\
Scalar spectral index (at $k_0=0.05$ Mpc$^{-1}$)\tablenotemark{c} 
                       &\ensuremath{n_s}     &\ensuremath{0.93}
                       &\ensuremath{0.03}       &\ensuremath{0.03}\\
Running index slope (at $k_0=0.05$ Mpc$^{-1}$)\tablenotemark{c}
                       &\ensuremath{dn_s/d\ln{k}}     &\ensuremath{-0.031}
                       &\ensuremath{0.016}       &\ensuremath{0.018}\\

Tensor-to-scalar ratio (at $k_0=0.002$ Mpc$^{-1}$) &\ensuremath{r}
                       &$<0.90$
                       &95\% CL
                       &---\\
Redshift of decoupling &\ensuremath{z_{dec}}     &\ensuremath{1089}
                       &\ensuremath{1}       &\ensuremath{1}  \\
Thickness of decoupling (FWHM) &\ensuremath{\Delta z_{dec}}     &\ensuremath{195}
                       &\ensuremath{2}       &\ensuremath{2}\\
Hubble constant        &\ensuremath{h}     &\ensuremath{0.71}
                       &\ensuremath{0.04}       &\ensuremath{0.03}\\
Age of universe (Gyr)  &\ensuremath{t_0}     &\ensuremath{13.7}
                       &\ensuremath{0.2}       &\ensuremath{0.2}\\
Age at decoupling (kyr) &\ensuremath{t_{dec}}     &\ensuremath{379}
                       &\ensuremath{8}       &\ensuremath{7}\\
Age at reionization (Myr, 95\% CL)) &\ensuremath{t_r}     
                          &180& 220 & 80 \\
Decoupling time interval (kyr) &\ensuremath{\Delta t_{dec}}     &\ensuremath{118}
                       &\ensuremath{3}       &\ensuremath{2}\\
Redshift of matter-energy equality &\ensuremath{z_{eq}}     &\ensuremath{3233}
                       &\ensuremath{194}       &\ensuremath{210}\\                     
Reionization optical depth &\ensuremath{\tau}     &\ensuremath{0.17}
                       &\ensuremath{0.04}       &\ensuremath{0.04}\\
Redshift of reionization (95\% CL) &\ensuremath{z_r} 
                         &20 & 10& 9\\
Sound horizon at decoupling ($^\circ$) 
                       &\ensuremath{\theta_A}    
                       &\ensuremath{0.598}
                       &\ensuremath{0.002}       
                       &\ensuremath{0.002}\\
Angular size distance (Gpc) &\ensuremath{d_A}     &\ensuremath{14.0}
                       &\ensuremath{0.2}       &\ensuremath{0.3}\\
Acoustic scale\tablenotemark{d}         &\ensuremath{\ell_A}     &\ensuremath{301}
                       &\ensuremath{1}       &\ensuremath{1}\\
Sound horizon at decoupling (Mpc)\tablenotemark{d} 
                       &\ensuremath{r_s}     &\ensuremath{147}
                       &\ensuremath{2}       &\ensuremath{2}\\
\enddata

\tablenotetext{a}{from {\sl COBE} \citep{mather/etal:1999}}
\tablenotetext{b}{derived from {\sl COBE} \citep{mather/etal:1999}}
\tablenotetext{c}{$l_{eff}\approx 700$}
\tablenotetext{d}{$\ensuremath{\ell_A}\equiv \pi \theta_A^{-1}$~~~
                  $\theta_A\equiv r_s\; d_a^{-1}$}
\end{deluxetable}

A power spectrum of primordial mass fluctuations with a scale invariant spectral index 
is given by $P(k) = A k^{n_s}$ with $n_s= d\ln P / d \ln k$. 
Inflationary models predict a running spectral index \citep{kosowsky/turner:1995}, and 
our best fit model uses a power spectrum of primordial mass fluctuations with a 
scale-dependent spectral index:
\begin{equation}
P(k) = P(k_0) \left( \frac{k}{k_0}\right)^{n_s(k_0) + (1/2) (d n_s/d \ln k) \ln (k/k_0)}.
\end{equation}
As in the scale-independent case, we define
\begin{equation}
n_s(k) \equiv  \frac{d \ln P}{d \ln k}, 
\end{equation}
so
\begin{equation}
n_s(k) = n_s(k_0) + \frac{d n_s}{d \ln k} \ln \left( \frac{k}{k_0} \right) 
\end{equation}
(with $d^2n_s/d \ln k^2 = 0$).
The definition for $n_s$ used here includes a factor of $(1/2)$ difference from the 
\citet{kosowsky/turner:1995} definition.  \citet{peiris/etal:2003} explore the
implications of this running spectral index for inflation.
The best fit values of $A$, $n_s$ and $dn_s/d \ln k$ are in Table \ref{tbl-2} for
$k_0 = 0.05$ Mpc$^{-1}$.  
$A$ is the normalization parameter in CMBFAST version 4.1 with 
option UNNORM.  The amplitude of curvature fluctuations at the horizon 
crossing is
$\left \vert \Delta_R(k_0) \right \vert^2 = 2.95 \times 10^{-9} A$.
We discuss the implications of the measured
values of these parameters in \S\ref{inflation} and in \citet{peiris/etal:2003}.

The \iMAP\ data constrains the properties of both the dark matter and the dark energy 
in the following ways:

\begin{enumerate}

\item[(a)]The \iMAP\ detection of reionization at $z \sim 20$ 
is incompatible with the presence of significant 
warm dark matter density.  
Since the warm dark matter moves too fast to cluster in small objects,
the first objects do not form in this scenario until $z \sim 8$  
\citep{barkana/haiman/ostriker:2001}.  

\item[(b)] The running spectral index implies a lower amplitude for mass
fluctuations on the dwarf galaxy scale.  Dark matter simulations of
models \citep{ricotti:2002} find that the dark matter mass profiles 
depend upon the spectral index on the relevant mass scale.  
Thus, the shallower spectral index implied by our best fit model
may solve the CDM dark matter halo profile problem 
\citep{moore/etal:1998, spergel/steinhardt:2000}.

\item[(c)] While the \iMAP\ data alone are compatible with a wide range of possible properties
for the dark energy, the combination of the \iMAP\ data with either the HST key
project measurement of $H_0$, the 2dFGRS measurements of the galaxy power spectrum
{\it or} the Type Ia supernova measurements requires that 
the dark energy be 73\% of the total density of
the Universe, and that the equation of state of the dark energy satisfy
\ensuremath{w < -0.78 (95\%\mbox{\ CL})}.

\end{enumerate}

\subsection{Best Fit Parameters}

\iMAP's measurements of the baryon density, Hubble constant, and
age of the universe strengthen the cosmic consistency that underlies the Big Bang model.

{\sl ATOMIC DENSITY (}$\Omega_b h^2${\sl)}:
\iMAP\ measures the atomic density at recombination 
to an accuracy of 4\%
through the shape of the angular power
spectrum, and particularly through the ratio of the heights of the first to second
peak \citep{page/etal:2003c, spergel/etal:2003}.  Our best fit value is
\ensuremath{\Omega_bh^2 = 0.0224 \pm 0.0009}.
The baryon density is also probed via abundance measurements of [D]/[H] \citep{omeara/etal:2001,
pettini/bowen:2001, dodorico/dessauges-zavadsky/molaro:2001}. It is impressive that
$\Omega_b h^2$ is the same at $z=1089$ as measured via the CMB as it is at $z=10^9$ from
Big Bang nucleosynthesis.
Thus we find cosmic consistency of the baryon density throughout cosmic time and measurement
technique.

{\sl HUBBLE CONSTANT (}$H_0${\sl):}
The \iMAP\ measurements of the age and $\Omega_m h^2$ yield a measurement of 
$H_0=71^{+4}_{-3}$ km s$^{-1}$ Mpc$^{-1}$ that is
remarkably consistent with the HST Key Project value of $H_0 = 72\pm 3\pm 7$~km s$^{-1}$ 
Mpc$^{-1}$
\citep{freedman/etal:2001}, but with smaller uncertainty. Recent measurements of the
Hubble constant from gravitational lens timing and the Sunyaev-Zeldovich
effect yield independent estimates that are generally consistent, but with
larger uncertainties at present.  Through a variety of measurement techniques that
sample different cosmic times and distances we find cosmic consistency on $H_0$.

{\sl AGE OF THE UNIVERSE (}$t_0${\sl ):}
The first acoustic peak in the CMB power spectrum represents a known acoustic
size (\ensuremath{r_s = 147 \pm 2 \mbox{ Mpc}}) at a known redshift 
(\ensuremath{z_{dec} = 1089 \pm 1}).  
From these, \iMAP\ measures the age of the universe (\ensuremath{t_0 = 13.7 \pm 0.2 \mbox{ Gyr}})
to an accuracy of $\sim 1$\% 
by determining the CMB light travel time over the distance determined by the 
decoupling surface (\ensuremath{d_A = 14.0^{+ 0.2}_{- 0.3} \mbox{ Gpc}}) and the geometry of the 
universe (i.e., flat).  The age of the
universe is also estimated via stars in three ways:
\begin{enumerate}
\item[(i)] the main sequence turn-off in globular clusters yielding a cluster age of
$12 \pm 1$~Gyr \citep{reid:1997};
\item[(ii)] the temperature of the coldest white dwarfs in globular clusters yielding
a cluster age of $12.7 \pm 0.7$~Gyr \citep{hansen/etal:2002}, and
\item[(iii)] nucleosynthesis age dating yielding an age of $15.6 \pm 4.6$~Gyr
\citep{cowan/etal:1999}.
\end{enumerate}
These stellar ages are all consistent with age of the universe found by \iMAP.

{\sl MATTER DENSITY (}$ \Omega_m h^2${\sl ):}
The matter density affects the height and shape of the acoustic peaks.
The baryon-to-matter ratio determines
the amplitude of the acoustic wave signal and the matter-to-radiation ratio
determines the epoch, $z_{eq}$, when the energy density of matter equals the 
energy density of radiation.  The amplitude of the
early integrated Sachs-Wolfe effect signal is sensitive
to the matter-radiation equality epoch.
From these effects \iMAP\ measures the matter density, $\Omega_m h^2$,to an accuracy
of $\sim 5$\%.
Large scale structure observations measure $\Omega_m h$ through
the shape of the power spectrum.  When combined with estimates
of $h$, this yields $\Omega_m h^2$.
Large scale velocity field measurements yield $\Omega_m^{0.6} b^{-1}$, 
where $b$ is the bias in how the galaxy power spectrum traces the matter power spectrum 
($P_{\rm gal}=b^2 P(k)$). 
Galaxy bispectrum measurements yield $b$, allowing
for estimates of $\Omega_m$.  From the galaxy data, \citet{verde/etal:2002} find
$\Omega_m = 0.27 \pm 0.06$, which is consistent with the \iMAP\ result 
of \ensuremath{\Omega_m = 0.27 \pm 0.04}.

Cluster lensing observations yield measurements of the
total mass in the cluster.  X-ray measurements give both the
baryonic mass and the total mass through the gravitational
potential.  Sunyaev-Zeldovich effect observations give a
different determination of the baryonic mass in clusters.
The combined X-ray and SZ measurements give a
value of $\Omega_b \Omega_m^{-1} = 0.081^{+0.009}_{-0.011} h^{-1}$ 
\citep{grego/etal:2001}, which is reasonably consistent with
\ensuremath{0.17 \pm 0.01} 
from \iMAP.

\subsection{Implications for Inflation \label{inflation}}

\iMAP\ data tests several of the key predictions of the inflationary scenario 
(see \citet{peiris/etal:2003} for further discussion):

\begin{enumerate}

\item[(a)] Inflation predicts that the universe is flat.  As
noted in \S\ref{model} and discussed
in detail in \citet{spergel/etal:2003}, the combination of \iMAP\ data with
either $H_0$, Type Ia SNe, or large scale structure data
constrains $\vert 1-\Omega_{tot} \vert < 0.03$.

\item[(b)] Inflation predicts Gaussian random phase fluctuations.  \citet{komatsu/etal:2003}
shows that the CMB fluctuations have no detectable skewness and place strong constraints
on primordial non-Gaussianity.  \citet{komatsu/etal:2003}  also shows that 
the Minkowski functionals of the \iMAP\ data are consistent with the predictions of a 
model with Gaussian random phase fluctuations.

\item[(c)] Inflation predicts fluctuations on scales that appear to be super\-horizon scales 
in a Friedman-Robertson-Walker (FRW) cosmology. The \iMAP\ detection of an 
anti-correlation between polarization and temperature fluctuations on scales of 
$\sim 1^\circ - 2^\circ$ \citep{kogut/etal:2003}
confirms this prediction and rules out subhorizon causal mechanisms for generating 
CMB fluctuations
\citep{peiris/etal:2003}.

\item[(d)] Inflation predicts a nearly scale invariant spectrum of fluctuations, as seen by \iMAP.
\end{enumerate}

The \iMAP\ data, in combination with complementary cosmological data, 
not only test the basic ideas of the inflationary scenario but
also rule out broad classes of inflationary models, and therefore the data guide us
towards a specific workable inflationary scenario.  The \iMAP\ data place significant constraints
on $r$, the tensor-to-scalar ratio, $n_s$, the slope of the scalar fluctuations and
$dn_s/d \ln k$, the scale dependence of these fluctuations.  
The addition of an admixture of isocurvature modes does not improve 
the \iMAP\ model fits.

The best fit model to the combination of the \iMAP, ACBAR, CBI, 2dFGRS and the 
Lyman-$\alpha$ forest data has a spectral index that runs from $n > 1$ on the large scales
probed by \iMAP\ to $n < 1$ on the small scales probed by the 2dFGRS and the Lyman-$\alpha$
forest data.  Only a handful of inflationary models predict this behavior.  
The \citet{linde/riotto:1997} hybrid inflationary model is one example.  
The data, however, do not yet require $n>1$ on large 
scales: our best fit model has $n_s = 1.03 \pm 0.04$ at $k = 0.002$ Mpc$^{-1}$.

Our analysis of inflationary models \citep{peiris/etal:2003} marks the beginning of
precision experimental tests of specific inflationary models.  With the addition of 
on-going \iMAP\ observations and future improved analyses, \iMAP\ will be able to more 
accurately constrain $\tau$ and hence $n_s$ on large scales.   
When other CMB experiments are calibrated directly to the \iMAP\ sky maps, 
they will provide improved measurements of the temperature angular power 
spectrum for $l > 700$. The upcoming
release of the Sloan Digital Sky Survey (SDSS) power spectrum will provide an improved 
measurement of the galaxy power spectrum.  The SDSS Lyman-$\alpha$ forest data 
are expected to be a significant improvement over the data used in our current analysis.   
Looking further towards the future, ESA's {\sl Planck} mission will provide
improved measurements on the CMB angular power spectrum on smaller angular scales and 
should be able to improve constraints on $r$.  

\section{DATA PRODUCTS \label{data}}

All of the \iMAP\ data will be released.  In addition, several ancillary and analyzed 
data sets are released.  These include beam patterns, angular spectra, etc.  
Some software tools are also provided.  An Explanatory Supplement provides detailed information 
about the \iMAP\ in-flight operations and data products \citep{limon/etal:2003}.  
All \iMAP\ data products are distributed through the Legacy Archive for Microwave Background
Data Analysis (LAMBDA) at http://lambda.gsfc.nasa.gov.  This is a new NASA data center dedicated to the rapidly growing field
of microwave background data archiving and analysis.

\section{SUMMARY AND CONCLUSIONS}

(1)  \iMAP\ has produced high quality full sky maps in five widely separated
frequency bands.  These maps can be used to test cosmological models and 
serve as the primary legacy of the mission.

(2)  We have characterized and placed stringent limits on systematic measurement 
errors.  The calibration is based on the modulation of the CMB dipole, and is accurate 
to better than 0.5\%.

(3)  We have demonstrated the ability to separate the CMB anisotropy from Galactic 
and extragalactic foregrounds.  We provide masks for this purpose.  In addition, we 
have produced CMB maps where the Galactic signal is minimized.

(4)  We have a new determination of the dipole.  It is 
$3.346\pm 0.017$ mK in the direction 
($l$, $b$)$=$ ($263\ddeg 85\pm0 \ddeg 1$, $48\ddeg 25\pm0\ddeg 04$).

(5) We have a new determination of the quadrupole amplitude.  It is 
$Q_{rms}=8\pm 2\; \mu$K or $\Delta T_2^2=154\pm70$ $\mu$K$^2$.

(6)  We have placed tight new limits on non-Gaussianity
of the CMB anisotropy.
The coupling coefficient of a quadratic non-Gaussian term is limited to 
$-58 < f_{NL} < 134$ (95\% CL) \citep{komatsu/etal:2003}.

(7)  We have produced an angular power spectrum of the anisotropy with unprecedented 
accuracy and precision.  The power spectrum is cosmic variance limited for 
$l<354$ with a signal-to-noise ratio $>\!\!\!1$ per mode to $l=658$. 

(8)  We have, for the first time, observed the angular power spectrum of TE 
temperature-polarization correlations 
with sufficient accuracy and precision to place meaningful limits on cosmology.

(9)  We have detected the epoch of reionization with an optical depth of 
\ensuremath{\tau = 0.17 \pm 0.04}.  This implies a reionization epoch of  
$t_r=180^{+220}_{-80}$ Myr (95\% CL) after the Big Bang at a redshift of 
$z_r=20^{+10}_{-9}$ (95\% CL) for a range of ionization scenarios.  
This early reionization is incompatible with the presence of a significant 
warm dark matter density.

(10) We have fit cosmological parameters to the data.  We find results that are consistent 
with the Big Bang theory and inflation.  We find that the addition of a running spectral
index, while not required,  
improves the fit at the $\sim 2\sigma$ level.  
We provide values and uncertainties for a host of parameters 
based on this non-power-law inflationary model.  Our ``best'' values for cosmic parameters 
are given in Table \ref{tbl-2}.

(11) \iMAP\ continues to collect data and is currently approved for 4 years of 
operations at L$_2$.  The additional data, and more elaborate analyses, will help to 
further constrain models.  The addition of other continuously improving CMB and 
large scale structure observations is essential for progress towards the ultimate 
goal of a complete understanding of the global properties of the universe.

\acknowledgements

The \iMAP\ mission is made possible by the support of the Office of Space 
Sciences at NASA Headquarters and by the hard and capable work of scores of 
scientists, engineers, technicians, machinists, data analysts, budget analysts, 
managers, administrative staff, and reviewers.  We are grateful to the National Radio 
Astronomy Observatory, which designed and produced the HEMT amplifiers that made \iMAP\ 
possible.  We are grateful to A. Riess for providing the likelihood surfaces for the 
supernova data.  D. Finkbeiner supplied us with his full sky composite map of H$\alpha$ 
emission in advance of publication.  LV is supported by NASA through a Chandra Fellowship 
issued by the Chandra X-ray Observatory Center, operated by the Smithsonian Astrophysical 
Observatory. ML and GT are supported by the National Research Council.

\clearpage

\clearpage
\clearpage

\clearpage

\clearpage

\clearpage

\clearpage

\clearpage

\clearpage

\clearpage

\clearpage

\clearpage

\clearpage

\clearpage

\end{document}